\documentclass[12pt]{iopart}

\usepackage{gensymb}
\usepackage{braket}
\usepackage{graphicx}
\usepackage{amssymb}
\usepackage{bm}

\begin{document}

\title[Superconducting surface trap chips for microwave-driven trapped ions]{Superconducting surface trap chips for microwave-driven trapped ions}

\author{Yuta Tsuchimoto$^{1,\dag}$, Ippei Nakamura$^1$, Shotaro Shirai$^{1,2}$, \& Atsushi Noguchi$^{1,2,3,\ddag}$}

\address{
$^1$ Komaba Institute for Science (KIS), The University of Tokyo, Meguro-ku, Tokyo, 153-8902, Japan

$^2$ RIKEN Center for Quantum Computing (RQC), Wako, Saitama 351–0198, Japan

$^3$ Inamori Research Institute for Science (InaRIS), Kyoto-shi, Kyoto 600-8411, Japan
}

\ead{tyuta@g.ecc.u-tokyo.ac.jp$^{\dag}$ ,
u-atsushi@g.ecc.u-tokyo.ac.jp$^{\ddag}$}
\vspace{10pt}
\begin{indented}
\item[]July 2024
\end{indented}

\begin{abstract}
Microwave-driven trapped ion logic gates offer a promising avenue for advancing beyond laser-based logic operations. In future microwave-based operations, however, the joule heat produced by large microwave currents flowing through narrow microwave electrodes would potentially hinder improvements in gate speed and fidelity. Moreover, scalability, particularly in cryogenic trapped ion systems, is impeded by the excessive joule heat. To address these challenges, we present a novel approach: superconducting surface trap chips that integrate high-$Q$ microwave resonators with large current capacities. Utilizing sub-ampere microwave currents in superconducting Nb resonators, we generate substantial magnetic field gradients with significantly reduced losses compared to conventional metal chips. By harnessing the high $Q$ factors of superconducting resonators, we propose a power-efficient two-qubit gate scheme capable of achieving a sub-milliwatt external microwave input power at a gate Rabi frequency of 1 kHz.
\end{abstract}

%
% Uncomment for keywords
%\vspace{2pc}
%\noindent{\it Keywords}: XXXXXX, YYYYYYYY, ZZZZZZZZZ
%
% Uncomment for Submitted to journal title message
%\submitto{\JPA}
%
% Uncomment if a separate title page is required
%\maketitle
% 
% For two-column output uncomment the next line and choose [10pt] rather than [12pt] in the \documentclass declaration
%\ioptwocol
%

\section{Introduction}

Trapped ion systems are among the most promising platforms for quantum information processing \cite{Bruzewicz2019, pino2021demonst, Racetrack}. Hyperfine qubits realized in trapped ions enable high-fidelity single- and two-qubit gate operations, as well as long coherence times \cite{Ballance2016, Gaebler2016, schafer2018, Clark2021}. Universal control of those qubits has traditionally been achieved through the utilization of optical Raman transitions with carefully controlled multiple laser beams. Alternatively, laser-free entangling gates have been recently demonstrated using DC \cite{Mintert2001, Weidt2016, doi:10.1126/sciadv.1601540}, RF \cite{srinivas2021} or microwave magnetic field gradients \cite{ospelkaus2011, Harty2016, hHahn2019, Zarantonello2019, weber2024_2}. Notably, microwave-based approaches have achieved single- and two-qubit gates with low error rates of approximately $10^{-6}$ \cite{Harty2014, Leu2023} and $\lesssim 10^{-2}$ \cite{Harty2016, hHahn2019, Zarantonello2019, weber2024_2}, respectively, while the speed of two-qubit gates is about 150 $\mu$s \cite{weber2024_2}. These numbers approach the performance of typical laser-based gate schemes. The direct access to the hyperfine transitions enabled by magnetic field gradients provides high-fidelity operations that are not limited by spontaneous optical Raman emission. Moreover, utilizing mature microwave off-the-shelf components and on-chip microwave circuitry potentially enhances scalability due to their compact sizes and superior controllability of frequencies, phases, and amplitudes. 

A challenge encountered in microwave-based entangling operations is the Joule loss of sub-ampere AC current flowing in narrow on-chip microwave waveguides. Considering standard gold waveguides with a width of $10$ $\mu$m and a resistivity of $1$ n$\ohm$ m at a cryogenic temperature \cite{weber2024}, an AC current of 1 A at 1 GHz generates a power dissipation of $\sim 200$ mW per 1 mm length of a single trap segment. This substantial amount of Joule heat can induce changes in the local dielectric environment due to temperature rise or, in the worst case, cause irreversible damage to the trap chip \cite{Hahn2019multi}. These effects would limit gate speed and fidelity. Furthermore, especially for cryogenic operations to suppress anomalous heating \cite{Chiaverini2014, Brownnutt2015}, the excessive Joule heat would constrain the number of entangling gates that can be executed in parallel due to the limited cooling power.

To address these issues, we develop superconducting surface trap chips that incorporate high-$Q$ and large-current-capable superconducting microwave resonators. The low-loss property of superconductivity minimizes on-chip Joule heating, while its large current capacity enables the generation of large magnetic field gradients. In terms of the current capacity, the microwave critical currents of superconductors are governed by the theoretical limit known as the superheating field $H_{\rm sh}$, which represents the maximum magnetic field that can be sustained on the surface without triggering a transition from the Meissner state to the vortex state. The superheating field exceeds the lower critical field $H_{c1}$ by retaining the Meissner state as a metastable state. Recent advancements in superconducting materials, particularly in Nb, have achieved the critical field close to the superheating field, approximately 200 mT \cite{Marie2016, Padamsee2017}. This achievement has enabled the development of superconducting radio frequency cavities with enhanced high-current capacity and significantly increased $Q$ factors of $\sim 10^{10}$ \cite{Gurevich2006}.

%For sustaining large currents, niobium is a highly promising candidate due to its high critical field. In the field of particle accelerators, extensive efforts have focused on enhancing the critical field of Nb against RF fields through material and design improvements \cite{Padamsee2017}. Recent advancements have brought the critical field close to its theoretical limit, the so-called superheating critical field, of approximately 200 mT \cite{Marie2016}. Consequently, superconducting radio frequency (SRF) cavities with increased Q factors, reaching up to $\sim 10^{10}$, and improved high-current capacity have been realized \cite{Gurevich2006}.

%Motivated by the high-$Q$ and large-current capacity of bulk Nb cavities, we investigate the potential of superconducting Nb chips for microwave-driven trapped ion systems. 

Motivated by the high-$Q$ and large-current capacity of bulk Nb cavities, we explore the potential of Nb films to achieve large microwave currents and high-$Q$ factors when integrated into the surface trap chips. The structure of this paper is as follows. In Section 2, we embark on an investigation into the current-carrying capacity and $Q$ factors of superconducting surface trap chips. We begin by examining a fundamental structure, coplanar waveguide resonators, as a precursor to understanding the intricate dependencies of more complex chip designs. We fabricate high-$Q$ superconducting coplanar resonators and conduct a detailed examination of the underlying physics that limits the maximum current. Subsequently, we extend our investigation by fabricating a superconducting surface trap chip, which imposes additional geometric constraints. This step allows us to assess the feasibility of accommodating a large current, hence a large magnetic field gradient within the context of an actual surface trap geometry. In section 3, instead of using a typical single-sided surface trap chip, we propose to use a 3D-integrated superconducting trap geometry to enhance the microwave-field gradient. We fabricate such a 3D trap chip using a flip-chip assembly technique and characterize the maximum currents and the $Q$ factors at various temperatures. Moreover, we explore the response of the $Q$ factors to continuous and pulsed light illumination in order to emulate the conditions encountered during laser cooling and readout operations. In section 4, we discuss a power-efficient two-qubit gate scheme that benefits from the high $Q$ factor of superconducting microwave resonators. Finally, in section 5, we summarize the issues to be addressed for further improvements to generate an even larger magnetic field gradient.

\section{Microwave characterization for superconducting surface trap chips}

\subsection{Superconducting coplanar waveguides}

\begin{figure}
    \centering
    \includegraphics[width=1\linewidth]{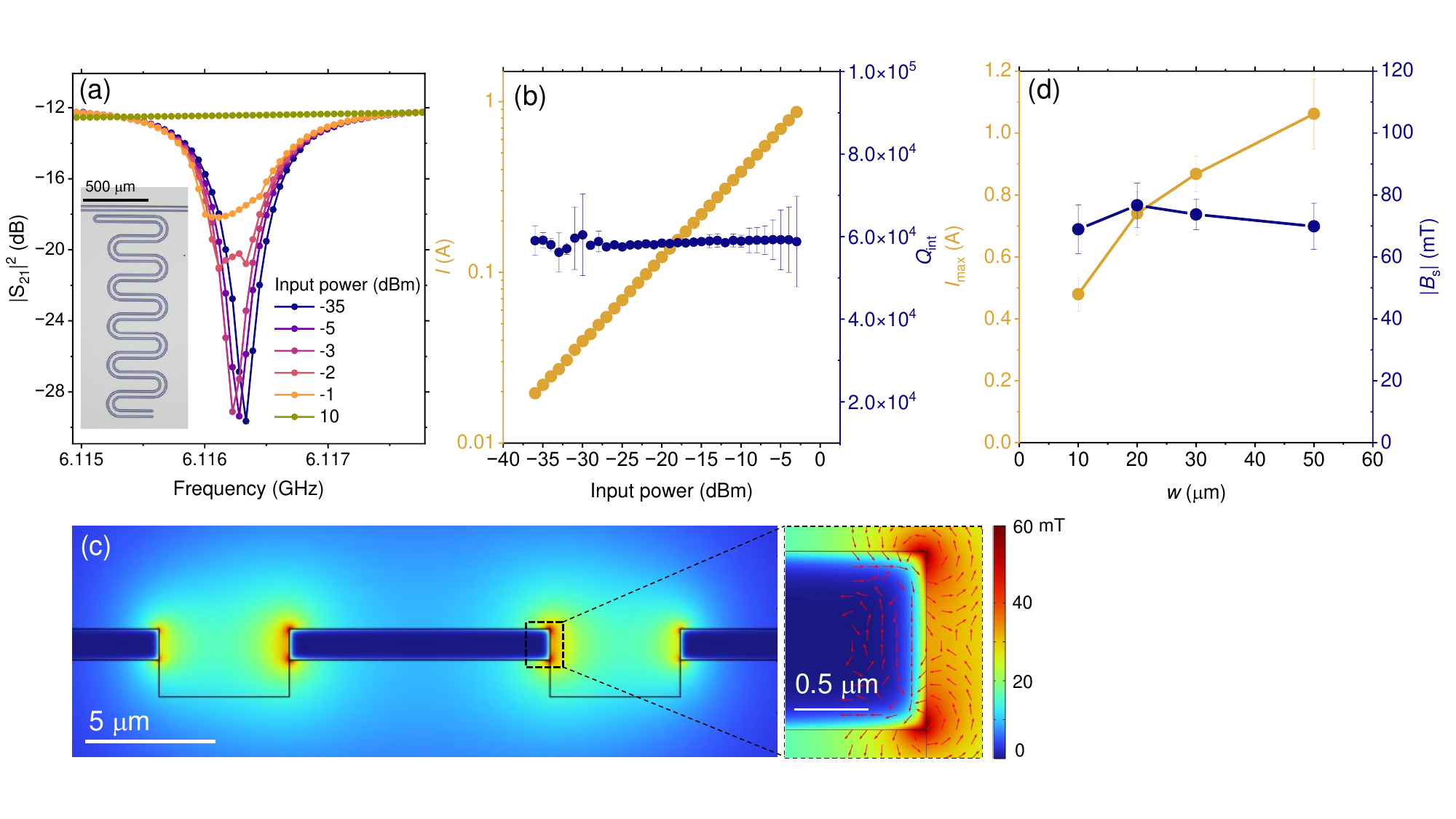}
    \caption{(a) $|S_{21}|^2$ spectra of a $\lambda/4$ coplanar waveguide resonator coupled to a transmission line, obtained at various input microwave powers. The film is made of 1.2 $\mu$m-thick superconducting Nb. The measurements were conducted at a temperature of 2 K. The inset displays an optical image of the studied resonator, featuring dimensions of a width $w = 30$ $\mu$m, a gap $s = 15$ $\mu$m, and a thickness $t = 1.2$ $\mu$m. (b) shows the dependence of the resonator's intra-current $I$ and internal quality factor $Q_{\rm int}$ on input power. The values of $I$ were derived from a calculated characteristic impedance and resonator mean photon numbers $\bar{n}_{\rm m}$ obtained by fitting $S_{21}$ in the complex plane. (c) A simulated magnetic field distribution across a cross-section of a Nb coplanar waveguide with dimensions of $w = 10$ $\mu$m, $s = 5$ $\mu$m, and $t = 1.2$ $\mu$m. The inset provides an enlarged view focusing on the right edge of the central line. The red arrows represent the normalized local magnetic field vectors around the corners. The simulation was performed using COMSOL by incorporating the properties of a superconducting Nb film, characterized by a complex permittivity as described in the main text. The simulation assumes a current of $I = 1$ A flowing through the central electrode at a frequency of 1 GHz. (d) Maximum resonator currents $I_{\rm max}$ for various values of $w$, with $s = 1/2 w$, alongside simulated surface magnetic field strengths $B_{s}$ at a corner when $I_{\rm max}$ flow through the central line. Error bars in the plots represent fitting uncertainties.}
    \label{fig:Fig1}
\end{figure}

First, we fabricated superconducting coplanar waveguide resonators to investigate how large microwave currents can flow in narrow superconducting electrodes. We sputtered 1.2 $\mu$m-thick Nb film on a high resistive Silicon substrate ($\sim$ 20 k$\Omega$ cm) immediately after removing the native oxide layer with diluted hydrogen fluoride solvent. Subsequently, we patterned the circuits by reactive ion etching with a photoresist mask.
The inset of Figure \ref{fig:Fig1}(a) displays a fabricated coplanar waveguide resonator with a resonance frequency of 6.116 GHz, featuring a width and gap of 30 $\mu$m and 15 $\mu$m, respectively. We cooled the sample down to 2 K using a Gifford-McMahon cryocooler.

We obtained the resonance spectrum of the resonator by measuring microwave transmission with a vector network analyzer through a feedline capacitively coupled to the resonator. Figure \ref{fig:Fig1}(a) shows the resonance spectra measured with various input microwave powers. As the input power increases, the resonance peak experiences a slight shift towards shorter frequencies within the linewidth. This small resonance shift is indicative of nonlinear kinetic inductance effect \cite{Thomas2020}. In addition to the shift, the spectra display a sudden reduction in resonance depth when the input power surpasses a critical value, causing the Lorentzian lineshape to transition into a flat-bottom shape. This phenomenon is attributed to power-dependent nonlinear dissipation described in \cite{Thomas2020}, where the power-dependent $Q$ factor is represented by a power-law expression: $Q_{\rm nl} = Q_{\rm ext} (P_{\rm c}/P_{\rm nl})^r$. Here, $P_{\rm c}$ denotes the critical power threshold at which nonlinear behaviour emerges, while $P_{\rm nl}$ represents the power dissipated by the nonlinear process. The exponent $r$ is a rational number dependent on underlying physical mechanisms. The observed flat-bottom shapes in the plot agree well with the case for $r > 1$, where a local breakdown of superconductivity in a Nb thin-film resonator is attributed to the origin of the nonlinearity. The further details of the local breakdown in our resonators will be discussed later in this section.

By fitting $S_{\rm 21}$ spectra, we extracted the resonator's external ($Q_{\rm ext}$) and internal ($Q_{\rm int}$) quality factors as well as the intra-photon numbers ($\bar{\rm n}_{\rm m}$). For the fitting, we used the algorithm described in \cite{Probst2015} and restricted our analysis to spectra acquired below $P_{\rm c}$, where the Lorentzian line shape of the resonance begins to distort. Following the onset of nonlinearity, a significant increase in resonator loss precludes any substantial rise in $\bar{n}_{\rm m}$. From $\bar{n}_{\rm m}$, we estimated the amplitude of the intra-current $I$ in the coplanar waveguide resonator using a 3D finite element method (FEM) with COMSOL. The COMSOL simulation estimates the relation between $\bar{n}_{\rm m}$ and $I$, which scales as $I \propto \sqrt{\bar{n}_{\rm m}}$. Figure \ref{fig:Fig1}(b) shows $I$ and $Q_{\rm int}$ as a function of the input power. The current sublinearly increases with increasing the input power and takes the maximum value of $I_{\rm max} \sim$ 0.87 A at the threshold power. This large microwave current is accompanied by a high internal $Q$ factor of $Q_{\rm int} = 6\times10^4$, contrasting with the typical value for normal conductors ($Q_{\rm int} \sim 40$ \cite{weber2024}). Hereafter, we denote $Q_{\rm int}$ specifically at the point of achieving $I_{\rm max}$ as $Q^{\rm m}_{\rm int}$.

To further enhance $I_{\rm max}$, understanding the underlying origin of the nonlinear dissipation is imperative. The dissipation associated with the local breakdown is expected to be caused by, for instance, a surface magnetic field $B_{\rm s}$ exceeding the critical field $B_{\rm c}$ or the heating of impurities in a superconducting film due to a large AC current. Assuming that our nonlinear dissipative process originates from such local breakdown, we first examine the magnetic field distribution around the surface of a superconducting coplanar waveguide. To this end, we performed a cross-sectional FEM simulation using COMSOL. We model superconducting Nb electrodes by considering them as an environment with the following complex permittivity \cite{Niepce2020}:
\begin{equation} \label{eq:1}
\epsilon_r(\omega, T) = \epsilon_0 - \frac{1}{\omega^2\mu_0\lambda(T)^2} -i\frac{\sigma_1(\omega, T)}{\omega},  
\end{equation}
where $\epsilon_0$ and $\mu_0$ represent vacuum permittivity and permeability, respectively. The real part of the Mattis-Bardeen conductivity, $\sigma_1(\omega,T)$ has been set to zero to simulate loss-less superconductivity. In this way, we can take the magnetic penetration depth $\lambda(T)$ into account by considering the Maxwell-London equations at a given frequency. Here, we consider the case where $T \ll T_{\rm c}$ and assumed $\lambda(0) = 50$ nm \cite{maxfield1965superconducting}. We set the dielectric constant of the Silicon substrate to be $\epsilon_{\rm silicon}$ = 11.9 and the top-half calculation space as a vacuum. Figure \ref{fig:Fig1}(c) shows the simulated field distribution in the cross-section of a coplanar waveguide with $w = 10$ $\mu$m and $s = 1/2w$. The magnetic field strength inside Nb electrodes is exponentially suppressed over the distance of $\lambda = 50$ nm from the surface, as defined in equation (\ref{eq:1}). As expected from Lenz’s law, the field concentrates at the corners of the electrodes, where the superconductivity provably breaks down at first when one ramps the current up. 

To verify the hypothesis, we fabricated various resonators with different widths $w$ and gaps $s = 1/2w$ and estimated the maximum magnetic field $|B^{\rm max}_{\rm s}|$ at the corners when $I_{\rm max}$ was achieved. Figure \ref{fig:Fig1}(d) shows the measured $I_{\rm max}$ at different geometries: $w$ ranges from 10 to 50 $\mu$m, $s$ is fixed to be $s$ = 1/2$w$, and the thickness $t$ is 1.2 $\mu$m for all resonators. The maximum current $I_{\rm max}$ goes up from 0.48 to 1.1 A with increasing $w$ and $s$. We remark $Q_{\rm int}^{\rm m} \gtrsim 10^4$ for all resonators. We calculated $|B^{\rm max}_{\rm s}|$ at the corners by using COMSOL assuming these $I_{\rm max}$ flowing in the signal electrodes. The calculated $|B^{\rm max}_{\rm s}|$ at the corners are nearly constant around $\sim 70$ mT within the fitting error bars despite the current increase of more than a factor of two with increasing the width and gap. The constant breakdown field at the corners suggests that the concentrated field locally breaks the superconductivity at the corners first, which induces significant microwave loss to the system. This limits the amount of overall current flowing into the signal electrodes, leading to the determination of $I_{\rm max}$ in our coplanar geometry. However, the estimated magnitude of $|B^{\rm max}_{\rm s}|$ falls below the superheating field of $B_{\rm sh} \sim 200$ mT typically observed in bulk Nb microwave cavities utilized for particle accelerators \cite{Padamsee2017}. This discrepancy may stem from differences in material quality, notably the residual resistance ratio (RRR). While bulk Nb cavities regularly exhibit RRR values in the several hundred, our sputtered Nb films demonstrate a lower value of RRR $\sim 4$. The local heating of the material impurities at the corners seems to be the origin of the nonlinear dissipation in our superconducting resonators.

%The geometry dependence of the field distribution on a coplanar waveguide can be analytically calculated when $t <= \lambda$. The simple equation predicts reducing the field concentration by increasing $w$ and $s$. Here, to estimate the geometry dependence with a thick film coplanar waveguide more precisely, we again utilized COMSOL to 

\begin{figure}
    \centering
    \includegraphics[width=1\linewidth]{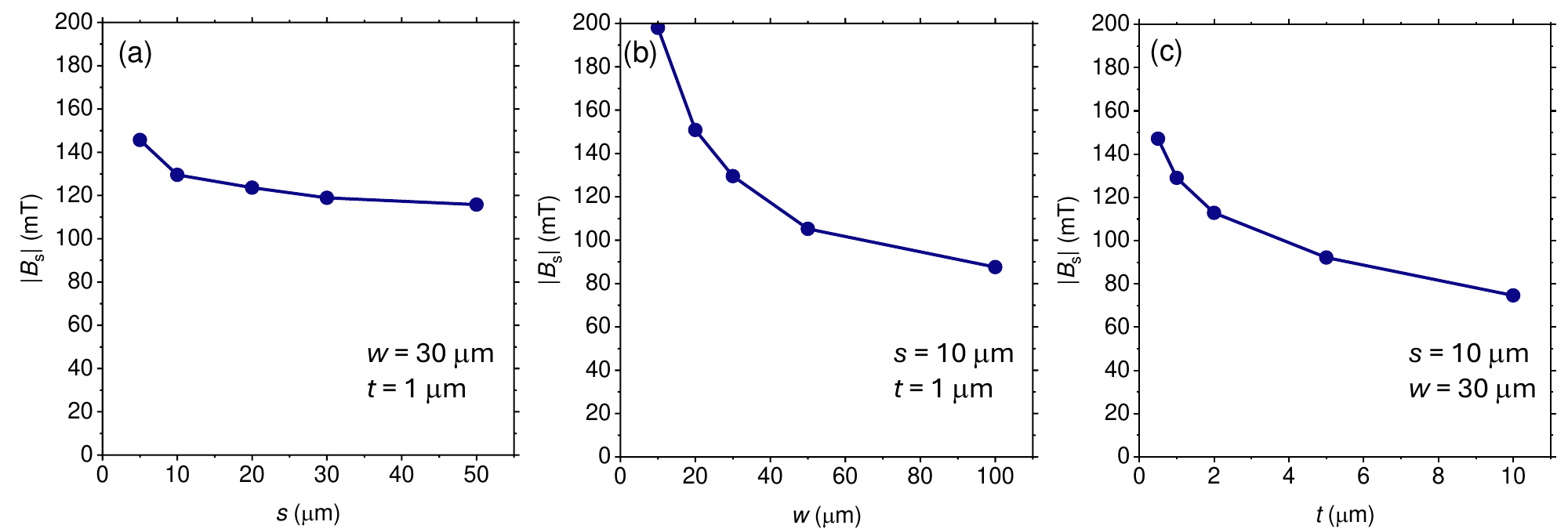}
    \caption{Geometry dependence of simulated $|B_{\rm s}|$ at a film corner by COMSOL, assuming $I = 1$ A flows in the central line of a Nb coplanar waveguide. (a), (b), and (c) display $s$, $w$, and $t$ dependencies, respectively. For each dependence, fixed dimensions are written in each figure.}
    \label{fig:Fig2}
\end{figure}

\subsection{Coplanar geometry dependence of magnetic field concentration}
Next, we simulate the influence of geometric parameters, specifically $w$, $s$, and $t$, on a surface magnetic field $|B_{\rm s}|$ at the electrode film corners. In contrast to the previous section where we applied maximum currents that break the superconductivity, here, we assume a constant current of $I = 1$ A flowing into the signal electrodes to investigate the geometric parameter dependence on $|B_{\rm s}|$. This provides insights crucial for optimizing trap chip design. Figure \ref{fig:Fig2} illustrates variations in $|B_{\rm s}|$ at a film corner for different geometric parameters. We observe a decrease in $|B_{\rm s}|$ as these parameters increase. In particular, the changes in $w$ and $t$ significantly impact $|B_{\rm s}|$ compared to the variations in $s$ within this parameter range. For typical surface trap chips \cite{weber2024, Allcock2013, hHahn2019, Hahn2019_2}, optimizing design involves narrowing $s$ ($s <~ 10$ $\mu$m) and increasing $t$ ($>~$ several $\mu$m) to minimize charge noise from dielectric surfaces. Typically, $w$ is several tens of micrometers due to geometric constraints for trapping ions at appropriate distances from the surface. To decrease $|B_{\rm s}|$ at the corners while satisfying these constraints, maximizing $t$ and designing $w$ to be as wide as possible within the limitations are effective strategies.

\subsection{Temperature dependence of $Q^{\rm m}_{\rm int}$ and $I_{\rm max}$}
\begin{figure}
    \centering
    \includegraphics[width=1\linewidth]{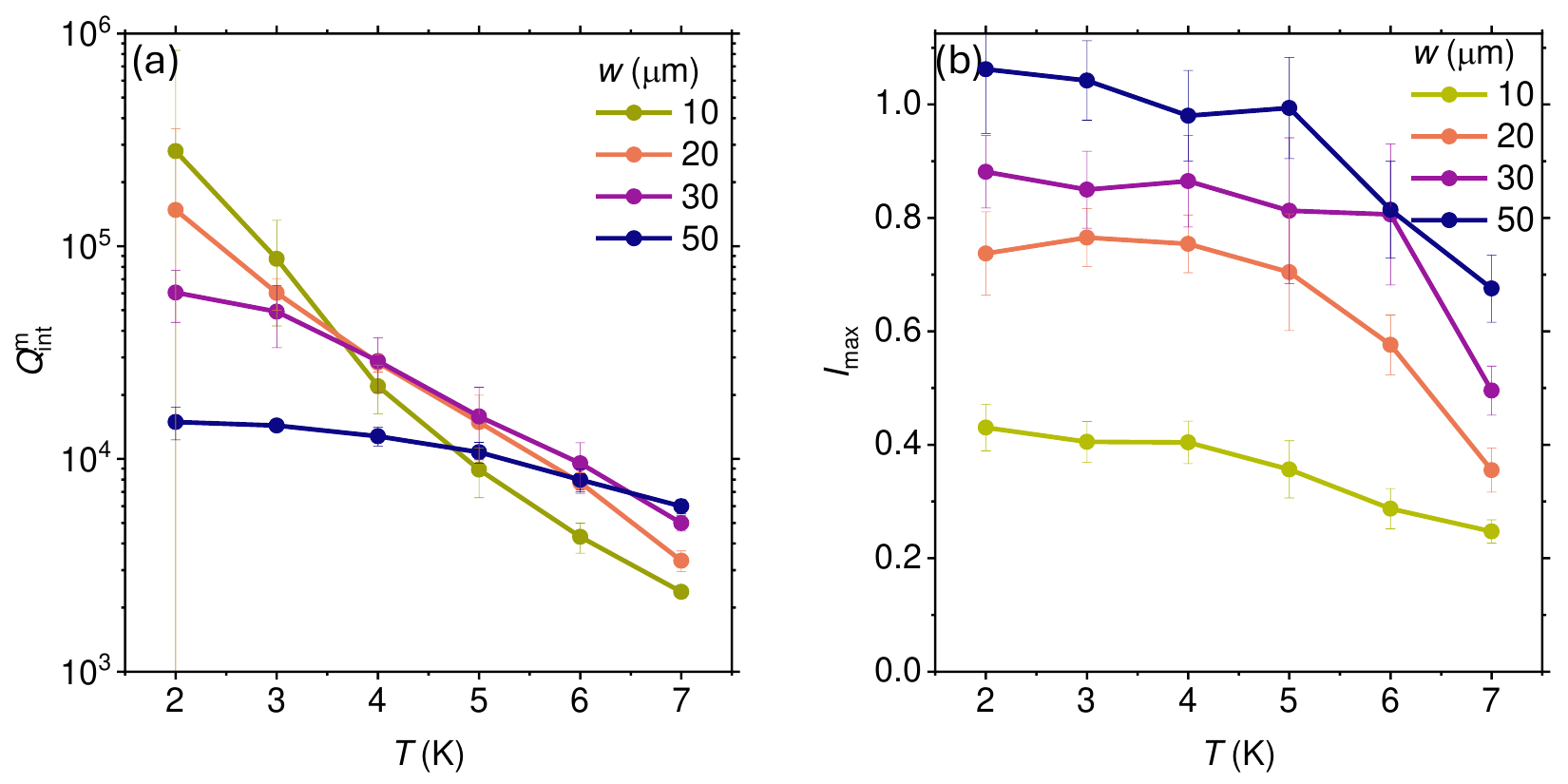}
    \caption{Temperature dependences of (a) internal quality factor $Q^{\rm m} _{\rm int}$ when (b) $I_{\rm max}$ flows in the resonator. The values of $Q^{\rm m} _{\rm int}$ and $I_{\rm max}$ were estimated from fitting $S_{\rm 21}$ measured at various temperatures. The dimensions of the measured samples are $w = 10, 20, 30, 50$ $\mu$m, $s = 1/2 w$, $t = 1.2$ $\mu$m. Error bars in the plots represent fitting uncertainties.}
    \label{fig:Fig3}
\end{figure}

We investigated the thermal tolerance of superconducting Nb resonators. By changing sample temperature using a heater in a cryostat, we assessed the temperature dependence of $I_{\rm max}$ and $Q_{\rm int}^{\rm m}$ for the resonators shown in Fig. \ref{fig:Fig1}(d). Figure \ref{fig:Fig3}(a) shows the temperature dependence of $Q_{\rm int}^{\rm m}$. At the lower temperatures, $Q_{\rm int}^{\rm m} $ is lower for the wider resonators, likely due to more effective coupling of electromagnetic fields to the bottom Copper holder compared to narrower resonators. As temperature increases from $T = 2$ to 5 K, $Q_{\rm int}^{\rm m} $ gradually reduces to $10^4$ across all resonators. This reduction is attributed to the generation of quasiparticles resulting from the absorption of thermal phonons. The higher current concentration at the corners of the narrower resonators may lead to increased microwave loss at higher temperatures compared to the wider resonators, thereby causing a steeper reduction in the $Q$ factors, as illustrated in the figure.  %Notably, the reduction of $Q^{\rm m} _{\rm int}$ is more significant for narrower coplanar resonators. The underlying cause of this trend is currently under investigation. 

Despite the decline in $Q_{\rm int}^{\rm m} $, $I_{\rm max}$ remains relatively stable up to $T = 5$ K for all resonators, as depicted in figure \ref{fig:Fig3}(b). This suggests that at these high $Q$ factors, $I_{\rm max}$, determined by $B^{\rm max}_{\rm s}$ at the film corners, remains unaffected by thermally excited quasiparticles.
%I_{\rm max}$ is not significantly limited by $Q_{\rm int}^{\rm m} $. 
%However, beyond $T = 6$ K, a gradual decline in $I_{\rm max}$ is observed, qualitatively consistent with the behavior described by $B_{\rm c1}(T) = B_{\rm c1}(0)(1-(T/T_{\rm c})^2)$ for $T \sim T_{\rm c} \sim 9$ K. 
Importantly, given that cryogenic ion trap systems readily achieve temperatures around $T \sim 5$ K \cite{Pagano_2019, 10.1063/5.0024423}, our findings suggest that such systems can preserve substantial current while maintaining low microwave loss.

\subsection{Superconducting surface trap chip}
\begin{figure}
    \centering
    \includegraphics[width=1\linewidth]{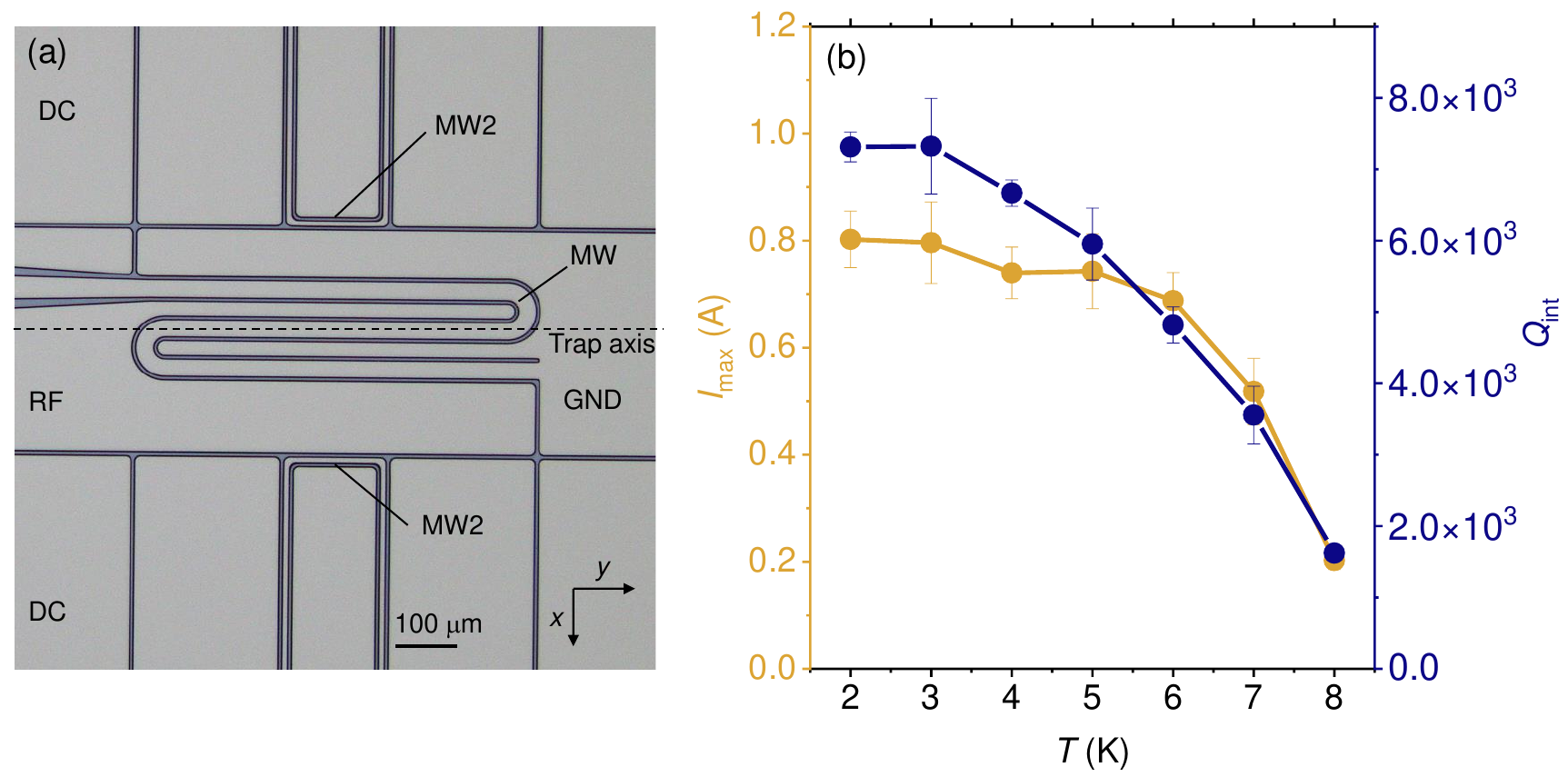}
    \caption{(a) An optical image of a fabricated Nb chip imitating a surface trap design demonstrated in \cite{hHahn2019}. The chip consists of a microwave resonator (denoted as MW) to create a magnetic field gradient, microwave waveguides (MW2) to create magnetic fields, RF, DC, and GND electrodes. The dashed line indicates the trap axis of the chip. (b) Estimated $I_{\rm max}$ and $Q^{\rm m}_{\rm int}$ by measuring $S_{\rm 21}$ of the microwave resonator at various temperatures. Error bars represent fitting uncertainties.}
    \label{fig:Fig4}
\end{figure}
So far, we have evaluated simple coplanar waveguide resonators to elucidate the geometry and temperature dependencies of $I_{\rm max}$ and $Q^{\rm m} _{\rm int}$. Here, we fabricate a superconducting surface trap chip including a microwave resonator as well as RF and DC electrode patterns to quantify $I_{\rm max}$ and $Q^{\rm m} _{\rm int}$ with a chip design that is more similar to an actual surface trap design. Figure \ref{fig:Fig4}(a) shows a fabricated trap chip of which the design is based on \cite{hHahn2019}. The sputtered Nb film thickness is 1.2 $\mu$m and the substrate is high-resistive silicon. The meander line (denoted as MW in figure \ref{fig:Fig4}(a)) is designed to be the current node of a $\lambda/4$ microwave resonator with a resonance frequency of 5.638 GHz. The nominal width of the meander is 30 $\mu$m and the gap to the ground is 5 $\mu$m. The phase relation between these three parallel lines creates a microwave field quadrupole, generating a field gradient with a field null point above the central line. The RF, DC electrodes and microwave lines (MW2) for homogeneous microwave magnetic fields are all floated. The other end of the microwave resonator (not shown) is capacitively coupled to a transmission feed line, which allows us to quantify $\bar{\rm n}_{\rm m}$ at a given input power as described before. In order to estimate the current amplitudes $I$ from the obtained $\bar{\rm n}_{\rm m}$, we performed a 3D COMSOL simulation of the chip and derived the relation between them, which scales as $I \propto \sqrt{\bar{\rm n}_{\rm m}}$. 

The estimated $I$ continuously increases with the input power until the nonlinearity appears. We extracted $I_{\rm max}$ and $Q_{\rm int}^{\rm m}$ and plotted them as a function of $T$ in figure \ref{fig:Fig4}(b). At $T = 2$ K, we observed a substantial current of $I_{\rm max} = 0.80$ A, slightly smaller than the value of $I_{\rm max} = 0.88$ A obtained for the coplanar waveguide resonator with $w = 30$ $\mu$m and $s = 15$ $\mu$m shown in figure \ref{fig:Fig1}(d). This small reduction can be understood as the effect of the narrower gap, which we discussed in figure \ref{fig:Fig2}(a). The obtained $I_{\rm max}$ remains nearly constant over $T$ of 2 - 5 K, consistent with the behavior shown in figure \ref{fig:Fig3}(b). We also demonstrated a high $Q_{\rm int}^{\rm m}$ of 7300 at 2 K and $Q_{\rm int}^{\rm m} = 6000$ at 5 K, indicating a superior quality factor compared to normal metal resonators. However, the $Q_{\rm int}^{\rm m}$ values are slightly lower than those of the superconducting coplanar resonators shown in figure \ref{fig:Fig3}(b). The reduced $Q_{\rm int}^{\rm m}$ may be attributed to a larger magnetic field component coupled to the bottom copper holder and free space due to the presence of the additional electrodes, which could function as antennas. The magnetic field gradient at the ion position is estimated to be $\partial{B_x}/\partial{z} = 30$ T/m for $I_{\rm max} = 0.80$ A using COMSOL. We remark that for practical use in ion trapping, the silicon substrate should be replaced with, for instance, a sapphire substrate to ensure a high withstand voltage for RF fields.

\section{Superconducting 3D Paul trap chip}
\begin{figure}
    \centering
    \includegraphics[width=1\linewidth]{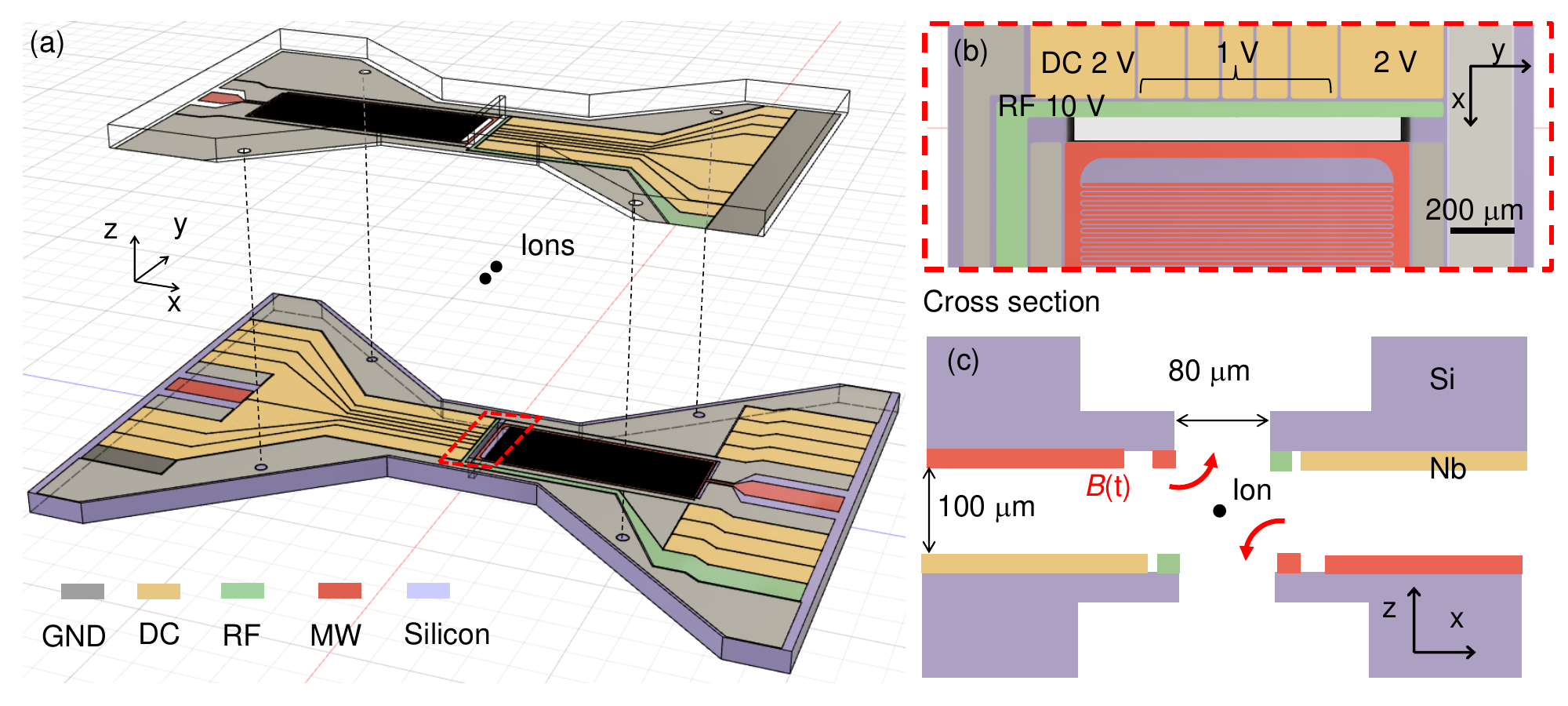}
    \caption{(a)Schematic representation of the 3D trap chip structure. The structure comprises top and bottom chips (Note: the depicted gap is not to scale), each featuring microwave (highlighted in red), RF (green), DC (yellow), and GND electrodes (grey) on a high-resistance silicon substrate (purple). A magnetic field gradient is established by two microwave LC resonators situated diagonally on the xz plane. Ions (depicted as black dots) are intended to be trapped between the chips. (b) Top view of the structure, with a focus on the microwave, RF, and DC electrodes within the region outlined by the dashed red rectangle in (a). The microwave resonator comprises an LC circuit, consisting of a U-shaped inductor and an interdigitated capacitor. (c) Cross-sectional schematic of the structure on the xz plane (not to scale). The electrodes are fabricated on the silicon substrates, facing each other between the top and bottom chips. Etching of the silicon substrates on the top and bottom sides of the ions allows for fluorescence collection and mitigates charge noise from silicon.}
    \label{fig:Fig5}
\end{figure}
In this section, we introduce a 3D trap design that offers a range of benefits, including a large field gradient and a deep trap potential. Figure \ref{fig:Fig5}(a) illustrates a schematic of our 3D superconducting Paul trap chip with electrodes situated on opposing faces of both the top and bottom chips. These chips are fabricated separately using standard microfabrication techniques and subsequently assembled using a flip-chip assembly (details provided later in this section). In the assembled chip, the superconducting RF and DC electrodes are positioned diagonally opposite each other on the chips, as depicted in the enlarged and cross-sectional schematics (figure \ref{fig:Fig5}(b) and (c)). Applying an RF voltage to the electrodes generates a trap potential that is more harmonic and deeper than 2D surface traps. This is expected to enhance trap lifetime and optimize ion transport. In addition, the segmented DC electrodes allow for axial confinement as well as compensation of ion position and motional polarization direction.

In order to generate a microwave magnetic field gradient, we incorporate two identical superconducting microwave LC resonators into diagonally opposite positions, oriented opposite to the DC and RF electrodes. Each LC resonator comprises a U-shaped inductor and an interdigitated capacitor (figure \ref{fig:Fig5}(b)). When two identical resonators exhibit electromagnetic coupling greater than their decay rates, the degenerate modes split into Helmholtz-like (H) and Anti-Helmholtz-like (AH) modes. These normal modes generate a homogeneous magnetic field and a magnetic field gradient at the ion position, respectively. The trap chip has horizontal optical accesses for cooling and readout laser beams together with a vertical through-hole window to collect ion fluorescence. A DC magnetic field can be applied in-plane to determine a spin quantization axis and resolve the frequency degeneracy of hyperfine transitions.

\subsection{RF trap potential}

We simulate a trap potential of our 3D trap. Here, we assume the gap between the top and bottom chips to be 100 $\mu$m and the width of the RF electrodes to be 50 $\mu$m. The horizontal distance between the RF and microwave electrodes is 80 $\mu$m, while the gap between the RF and neighbouring DC electrodes is 5 $\mu$m. The substrate assumed is insulating silicon with a dielectric constant of 11.9. The simulation considers a full potential \cite{Wesenberg2008}:
\begin{equation} \label{eq:2}
\Phi(\mathbf{r}) = \Phi_{\rm rf}(\mathbf{r}) + \Phi_{\rm dc}(\mathbf{r}),
\end{equation}
where a $^9$Be$^+$ ion is trapped by the combination of a DC potential $\Phi_{\rm dc}(\mathbf{r})$ and a RF pseudopotential $\Phi_{\rm rf}(\mathbf{r})$. In the adiabatic approximation, the RF pseudopotential experienced by an ion from an RF field is described by the ponderomotive potential \cite{Wesenberg2008}:
\begin{equation} \label{eq:3}
\Phi_{\rm rf}(\mathbf{r}) = \frac{Q^2}{4M\Omega_{\rm rf}^2}|E_{\rm rf}(\mathbf{r})|^2
\end{equation}
where $Q$ and $M$ are the charge and mass of a $^9$Be$^+$ ion, and $\Omega_{\rm rf}$ and $E_{\rm rf}(\mathbf{r})$ are the RF frequency and the simulated RF field amplitude. To obtain the pseudopotential, we performed a static 3D FEM simulation for $E_{\rm rf}(\mathbf{r})$ and calculated equation (\ref{eq:3}) with an RF frequency of $\Omega_{\rm rf} = 2\pi\times70$ MHz and an RF voltage of $V_{\rm rf} = 10$ V. We assumed the endcap DC bias to be 2 V and the bias of the inner DC electrodes to be 1 V  for both the top and bottom chips (see figure \ref{fig:Fig5}(b) for the bias configuration). Here, we bias the inner electrodes to lift the degeneracy of radial secular modes and to rotate the radial trap axes such that the propagation vector of a Doppler cooling laser beam has finite overlap with all motional axes.

\begin{figure}
    \centering
    \includegraphics[width=1\linewidth]{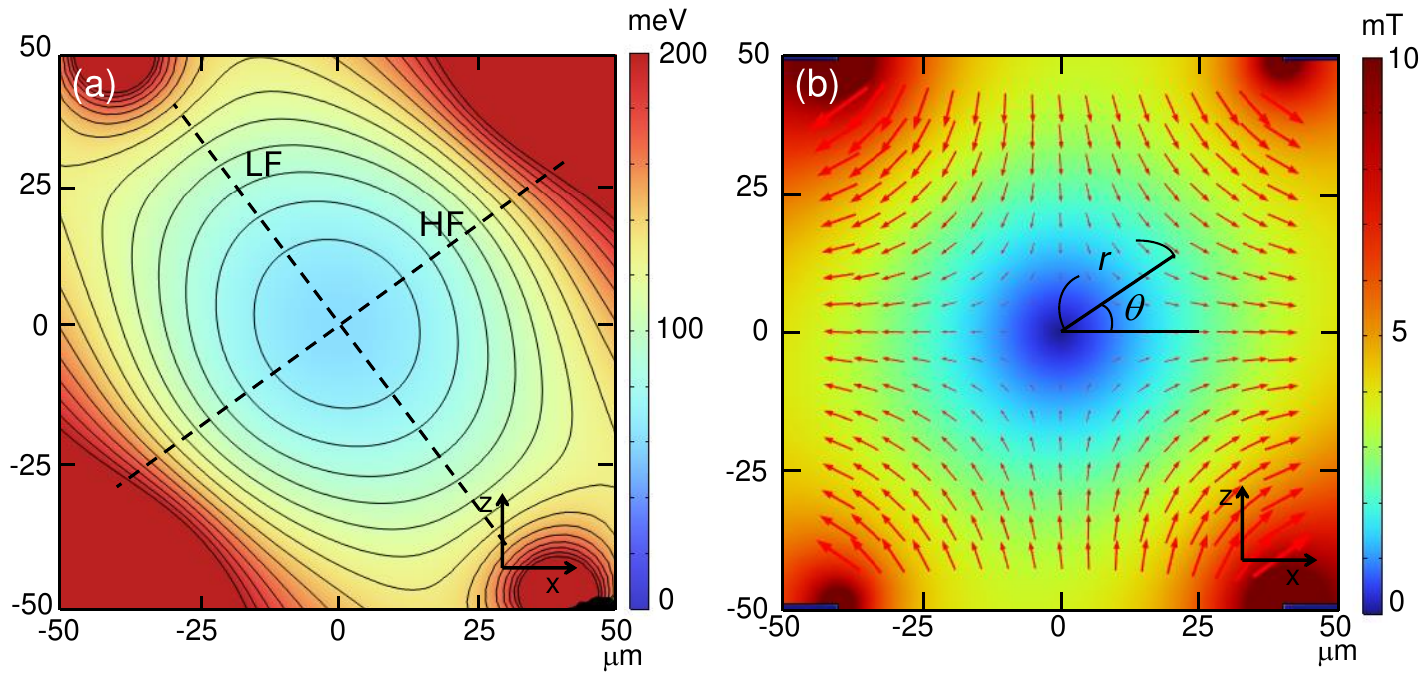}
    \caption{(a) Simulated effective RF trap potential $\Phi(\mathbf{r})$. The assumed voltages applied to the top and bottom electrodes are as follows: (RF electrodes, Endcap DC electrodes, Inner DC electrodes) = (10, 2, 1) V (See Fig.5(c) for the voltage configuration). The counter lines indicate $\Phi(\mathbf{r})$ by 10 meV step. The dashed lines show the low- (LF) and high-frequency (HF) axes. (b) Simulated magnetic field $|B^{\rm AH}(r,\theta)|$ distribution of the AH mode when $I = 1$ A flows in the microwave resonators. The red arrows represent the local vectors of $B(\mathbf{r})$. The angle from the x-axis is represented by $\theta$.}
    \label{fig:Fig6}
\end{figure}

Figure \ref{fig:Fig6}(a) displays the simulated $\Phi(\mathbf{r})$ on the radial plane. Due to the DC offset at the inner electrodes, the shape of the radial trap potential elongates diagonally, leading to the splitting of the radial mode into two different eigenfrequencies. By calculating the Hessian matrix, we determined the high and low secular frequencies to be ($\omega_{\rm HF}$, $\omega_{\rm LF}$) = $2\pi\times$(4.8, 4.5) MHz, with the corresponding radial trap axes positioned at ($\theta_{\rm HF}$, $\theta_{\rm LF}$) = (36, -53)$\degree$ from the x-axis, respectively. The shallowest trap depth is approximately 60 meV under the assumed RF and DC voltages. We estimated the $q$-parameter in the Mathieu differential equation \cite{Leibfried2003} to be $q_r \sim 0.19$ in the absence of DC offsets. Using the secular frequencies and the $q$ parameter, we derived $a$-parameters of approximately ($a_{\rm HF}$, $a_{\rm LF}$) $\sim$ (0.85, -1.7)$\times10^{-3}$ for the high and low frequency axes, respectively. Additionally, we estimated the axial trap potential to be around 70 meV, with a secular frequency of $\omega_{\rm ax} = 0.93$ MHz. This set of $a$- and $q$-parameters suggests that the trap condition lies deep within the stable regime \cite{Leibfried2003}.

\subsection{Microwave magnetic field}
Next, we examine the magnetic field distribution when the AH mode is excited. The field gradient generated by the AH mode can be used to induce spin-motion coupling for two-qubit gates. We conducted a 2D simulation with $I = 1$ A applied to the diagonally opposite microwave electrodes. Figure \ref{fig:Fig6}(b) illustrates the quadrupole magnetic field distribution $|B^{\rm AH}(r,\theta)|$ on the radial plane with arrows indicating the magnetic field direction. The magnetic null of the AH mode coincides with the radial center and overlaps with the RF null.
%The interaction between a spin transition of an ion and the magnetic field is given by $\mathbf{\mu} \times \textbf{\textit{B}}$ where $\mathbf{\mu}$ is the magnetic transition moment. The details of this product depend on experimental configuration. Here, we assume a weak DC magnetic field is applied along the xy plane with an angle $\phi$ from the x-axis. This magnetic field determines the spin quantization axis and resolves the frequency degeneracy of the hyperfine transitions. Then the inner product reads either $\mu_{\parallel}B_x(r,\theta)\cos{\phi}$ or $\mu_{\perp}B_z(r,\theta)/\sqrt{2}$ depending on whether the spin transition is $\pi$- or $\sigma$-transition, respectively. The factor of $1/\sqrt{2}$ for the $\sigma$-transition appears because only the left or right-circularly polarized component of the linearly polarized field interacts with the $\sigma$-transition. At the ion position of $r \sim 0$, the first derivative terms of the Tayler expansions represent spin-motion coupling: $\mu_{\parallel}\partial B_x(\theta)/\partial r \cos{\phi}$ or $\mu_{\perp}\partial B_z(\theta)/\partial r$.

The interaction between a spin transition of an ion and a microwave magnetic field from the AH mode is expressed as $\bm{\mu} \cdot \textbf{\textit{B}}^{\rm AH}$, where $\bm{\mu}$ denotes the magnetic dipole of an ion. The specific form of this product varies with experimental configurations. In this study, we consider a weak DC magnetic field $B_0$ applied in the xy plane at an angle $\phi$ relative to the x-axis. This magnetic field determines the spin quantization axis and resolves the frequency degeneracy of Hyperfine transitions. We also assume to drive a transition with $\Delta m = 0$, where $\Delta m$ represents the difference in the quantum number $m$ of the projection of the total atomic angular momentum $F$ along the quantization axis. Under these conditions, the amplitude of the inner product $\bm{\mu} \cdot \textbf{\textit{B}}^{\rm AH}$ takes the form of $\mu_{\parallel} B_x^{\rm AH}(r,\theta) \cos{\phi}$. Here, $\mu_{\parallel}$ represents the magnetic transition moments associated with the field components parallel to the quantization axis. At the ion's position of $r \sim 0$, the term arising from the first derivative in the Taylor series represents spin-motion coupling $\mu_{\parallel} \frac{\partial B_x^{\rm AH}(0, \theta)}{\partial r}q^0\cos{\phi}$ for a $\pi$-transition, where $q^0$ is the zero point motion of an ion.

The $\theta$ dependence of the magnetic field gradient along the radial direction can be determined by calculating the first derivative of the simulated field. The $B_x$ components of the field gradients along the HF and LF axes are estimated to be $\partial B_x^{\rm AH}(0, \theta_{\rm HF})/\partial r = 72$ T/m and $\partial B_x^{\rm AH}(0, \theta_{\rm LF})/\partial r = 56$ T/m, respectively. The maximum magnitude of the field gradient obtained with the 3D trap exceeds that of the 2D surface trap shown in figure \ref{fig:Fig4} by more than two-fold.

%In our geometry, the $B_x$ component of the numerically simulated $\theta$ dependence can be approximately expressed by $\partial B_x(\theta)/\partial r \sim \partial B_x(0)/\partial r \cos{\theta}$, where $\partial B_x(0)/\partial r = 90$ T/m for $I = 1$ A in each resonator. Here, $\partial B_x(0)/\partial r$ is linearly proportional to $I$. 

\subsection{3D trap chip fabrication}

\begin{figure}
    \centering
    \includegraphics[width=1\linewidth]{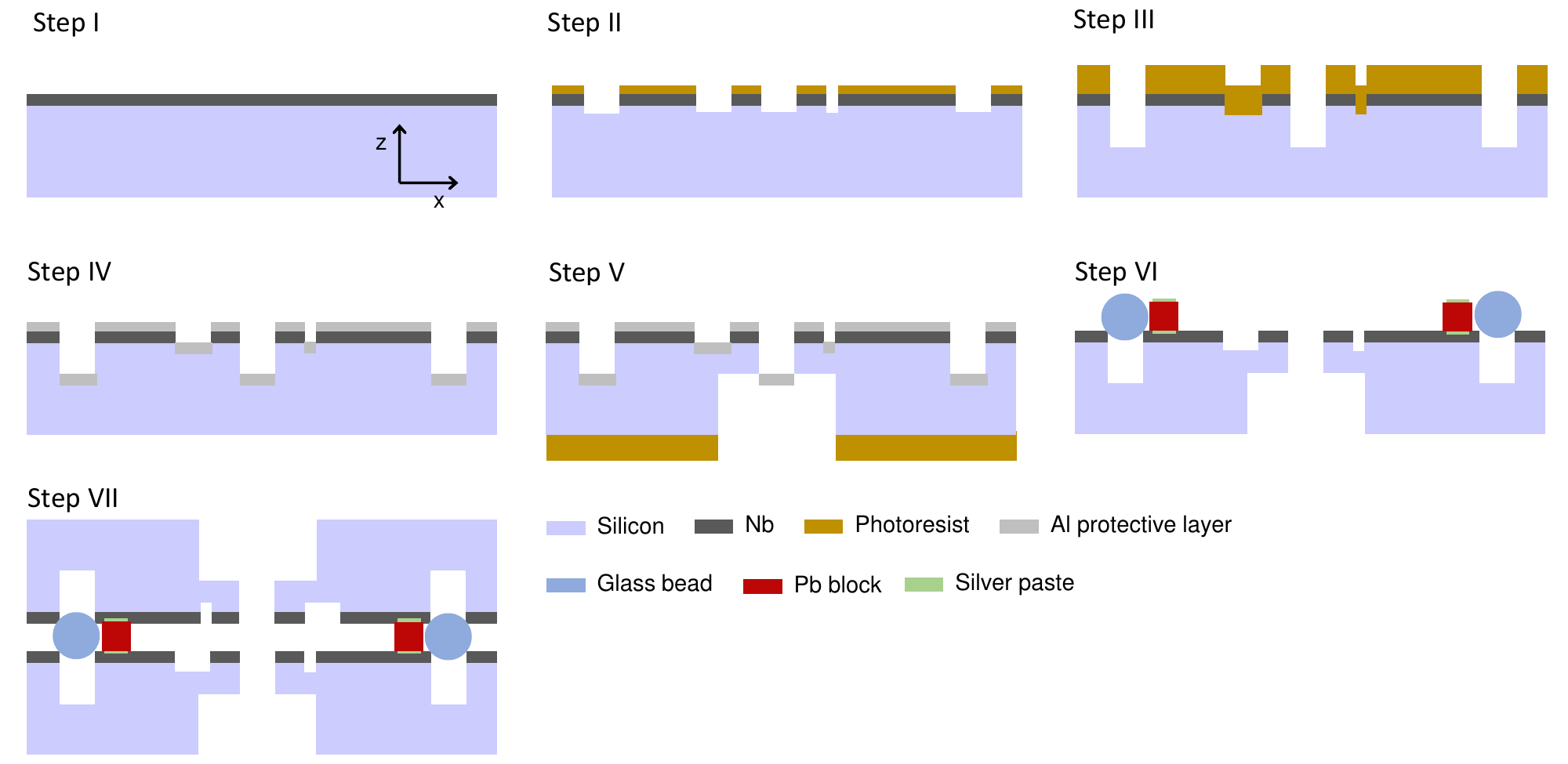}
    \caption{Schematic diagram of the fabrication process of the 3D trap chip (not to scale). The schematics show a cross-section along the $xz$ plane. Within this representation, some structures situated at different locations along the y-axis are projected onto the same xz plane. A Nb film is sputtered on top of a high-resistance Silicon wafer. The electrodes, the deep trenches, and the through holes are patterned by dry etching from the top and back sides of the wafer. Top and bottom chips processed in the same batch are assembled by a flip-chip technique with glass beads for precise alignment. Pb blocks with thin silver paste layers fix the top and bottom chips mechanically and establish electrical connections between them. The details of each step are described in the main text.}
    \label{fig:Fig7}
\end{figure}

We fabricated a 3D trap chip using typical microfabrication techniques (see figure \ref{fig:Fig7} for the processes). We first sputtered 1.2 $\mu$m-thick Nb film on a 300 $\mu$m-thick high-resistance silicon substrate (Step I). We patterned electrodes by photolithography followed by reactive ion etching (Step II). We defined the shape of the optical window, holes for chip assembly and outer chip boundary by etching Si, 100 $\mu$m deep from the top surface, using a Bosch process with a thick photoresist mask (Step III). After removing the mask, we deposited a 300 nm Aluminum layer to protect the surface from the following process (Step IV). We then created an optical window and detached the chip from the surrounding substrate by etching from the backside of the chip towards the patterns defined from the top surface (Step V). This etching was processed again by the Bosch process with a thick photoresist mask. After removing the resist mask and the Al protective layer, we measured resonance frequencies, maximum currents, and $Q$ factors for each resonator at cryogenic temperatures. We chose top and bottom resonators with a resonance frequency difference of 2 MHz and proceeded to the following flip-chip assembly. To assemble, we mounted glass beads with a diameter of 150 $\mu$m onto the alignment holes on the bottom chip. We also glued 100 $\mu$m-tall Pb blocks on the Nb pads with a small amount of silver paste (Step VI) to electrically connect the top and bottom chips. Using a manual die bonder, we flipped the top chip and pilled it on top of the bottom chip such that the glass beads on the bottom chip fit in the top alignment holes (Step VII). In this process, top Nb pads were also glued on top of the Pb blocks. The layers of silver paint were sintered at $180\degree$, and this sintering process mechanically fixed the top and bottom chips. With these glass beads and the alignment holes, we can align the chips with a horizontal misalignment of less than several $\mu$m and a tilt of less than 1 mrad. Finally, we mounted the assembled chip on a printed circuit board (PCB) and performed wire bonding from the PCB to the bottom Nb pads.

\subsection{Microwave characterization}
\begin{figure}
    \centering
    \includegraphics[width=1\linewidth]{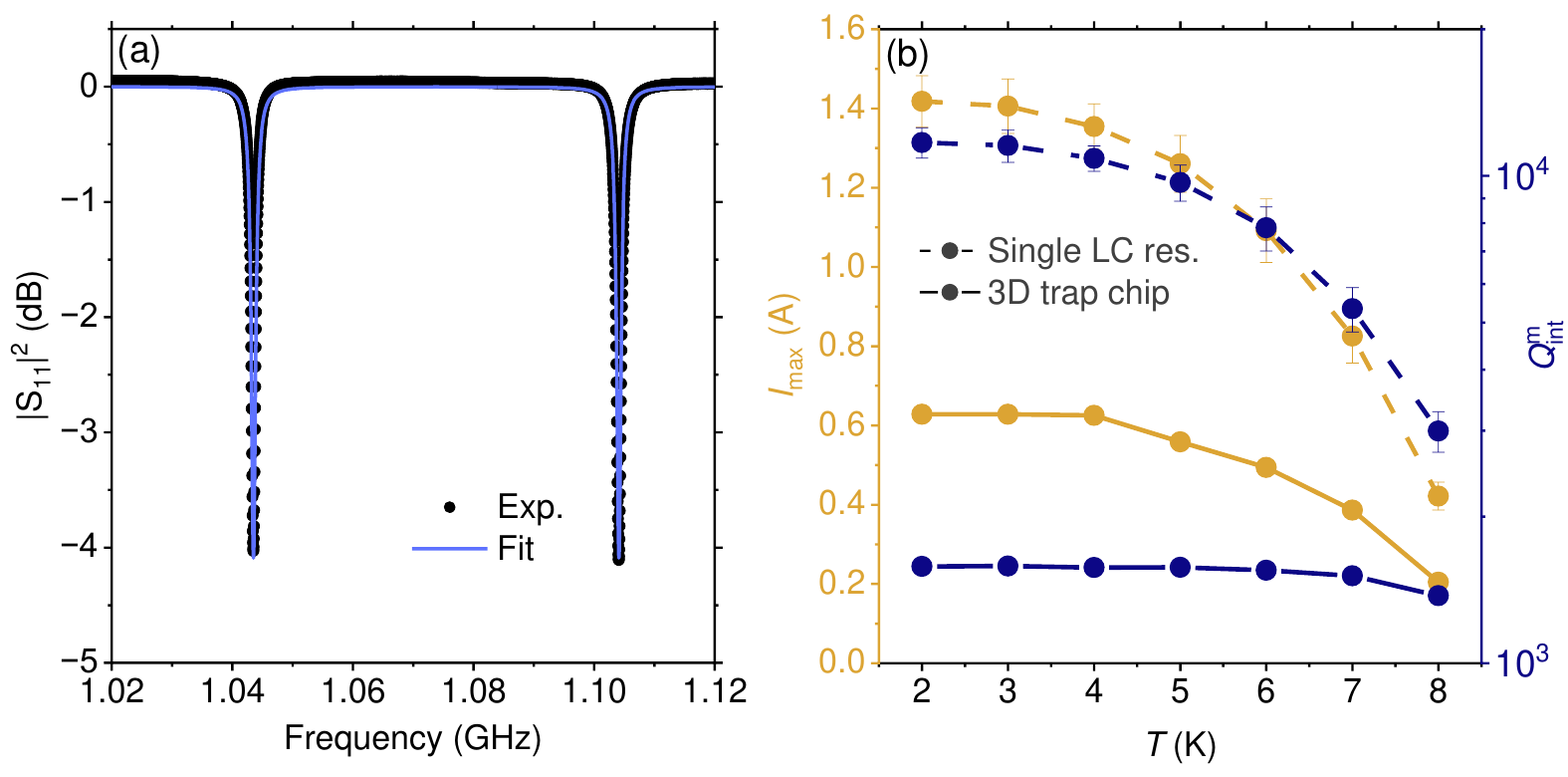}
    \caption{(a) Measured $|S_{\rm 11}|$ spectrum (black dots) of the coupled microwave resonators in a fabricated 3D trap chip. The coupled resonators were driven from Port1 depicted in Fig.5(a) and the reflected signal was fed to a VNA. The spectrum was measured at 2 K and normalized by a spectrum obtained above the superconducting transition temperature (9 K) to extract the response from the resonators. The blue line represents a fit to the spectrum modelled by coupled identical resonators. (b) The temperature dependence of $I_{\rm max}$ and $Q^{\rm m}_{\rm int}$ of the coupled resonator (solid lines) and a single LC resonator (dashed lines) without a through-hole for an optical window. The single LC resonator was coupled to a transmission line (notch-type configuration). The response of the single LC resonator was obtained through the transmission signal. Error bars represent fitting uncertainties.} 
    \label{fig:Fig8}
\end{figure}
In this section, we discuss the characteristics of the coupled microwave resonators in the 3D trap chip. To assess the resonators, we acquired $S_{\rm 11}$ spectra by sending microwave input signals to the bottom resonator (Port1) and then measuring returning signals using a VNA. The spectra were normalized by a spectrum obtained above the superconducting transition temperature to extract the response from the resonators. Figure \ref{fig:Fig8}(a) displays the $|S_{\rm 11}|^2$ spectrum of the coupled resonators at $T = 2$ K with an input power of 11.5 dBm. Two distinctive peaks at 1.044 and 1.104 GHz correspond to the AH and H modes, respectively. The observed normal mode splitting is $2g_{\rm m} = 2\pi\times60$ MHz, significantly larger than both the linewidths of each mode and the resonance frequency mismatch of $\delta_{\rm bt} = f_{\rm bottom} - f_{\rm top} = 2$ MHz that are measured before the assembly. In order to extract the resonator $Q$ factors, we fit the obtained spectrum using a model based on the input-output formalism for coupled resonators. The equations of motion for coupled resonators can be described as follows:
\begin{eqnarray}
\dot{a_t} = -(\frac{\omega_r}{2Q_{total}} -i \Delta) - i g_{\rm m} a_b - \sqrt{\frac{\omega_r}{Q_{\rm ext}}} a^{\rm in}_t\label{eq:4}\\
\dot{a_b} = -(\frac{\omega_r}{2Q_{total}} -i \Delta) - i g_{\rm m} a_t - \sqrt{\frac{\omega_r}{Q_{\rm ext}}} a^{\rm in}_{b},\label{eq:5}
\end{eqnarray}
with the photon input-output relations $a^{\rm out}_t = a^{\rm in}_t - \sqrt{\omega_r/Q_{\rm ext}} a_t$ and $a^{\rm out}_b = a^{\rm in}_b - \sqrt{\omega_r/Q_{\rm ext}} a_b$. Here, the subscripts $t$ and $b$ represent operators for the top and bottom resonators, respectively. The operators $a$, $a^{\rm in}$, and $a^{\rm out}$ denote the resonator photon annihilation operator, and the input and output operators, respectively. The detuning $\Delta$ expresses the frequency difference between the resonator resonance and a microwave drive. In this model, we neglect the small detuning $\delta_{\rm bt} \ll 2g_{\rm m}$ and assume that the top and bottom resonators have the same $Q_{\rm int}$ and $Q_{\rm ext}$. We solved equations (\ref{eq:4}) and (\ref{eq:5}) and fit the magnitude and phase components of $S_{\rm 11}$ spectrum together. Figure \ref{fig:Fig8}(a) shows that the fit spectrum (solid blue line) closely matches the experimental result, suggesting that the simple model assumed here reasonably describes the response of the actual coupled resonators.

Next, we estimated $Q_{\rm int}$, $Q_{\rm ext}$ and $\bar{n}_{\rm m}$ by measuring and fitting spectra at various temperatures and input powers. To estimate $I_{\rm max}$, we again acquired the relation between a current amplitude and $\bar{n}_{\rm m}$ using a 3D COMSOL simulation. In figure \ref{fig:Fig8}(b), we present the temperature dependence of $I_{\rm max}$ and $Q^{\rm m}_{\rm int}$ for the 3D trap (solid lines). The extracted $Q^{\rm m}_{\rm int}$ is constrained to approximately 1600 and exhibits no discernible temperature dependence up to 7 K. This is primarily attributable to the predominant microwave loss stemming from the large resistivity of the thin silver paint layers. An additional measurement indicated that the conductivity of the silver paint may approximate $10^3$ S/m at 2 K. According to COMSOL simulations, this low conductivity imposes limitations on the resonator's $Q$ factor, reaching approximately 1000.

On the other hand, the maximum current at 2 K was estimated to be $I_{\rm max} = 0.63$ A, generating $\partial B_x(\theta_{\rm HF})/\partial x = 48$ T/m at the AH mode frequency. The fabricated superconducting 3D trap chip shows a large magnetic field gradient and an order of magnitude low loss compared with normal conductor chips. Nevertheless, the fabricated superconducting 3D trap chip offers scope for improvement when contrasted with superconducting 2D resonators. A fabricated 2D LC resonator, employing the same resonator design as the 3D trap, demonstrates higher $I_{\rm max}$ with even lower loss over the entire temperature range (e.g., $I_{\rm max} = 1.4$ A and $Q^{\rm m}_{\rm int} = 1.1\times10^4$ at 2 K, see dashed lines). While the cause of the relatively low $I_{\rm max}$ for the 3D trap is currently under investigation, excessive heat generation resulting from the relatively low $Q_{\rm int}$ may contribute to the breakdown of superconductivity under strong driving conditions.

\subsection{Light sensitivity}
\begin{figure}
    \centering
    \includegraphics[width=1\linewidth]{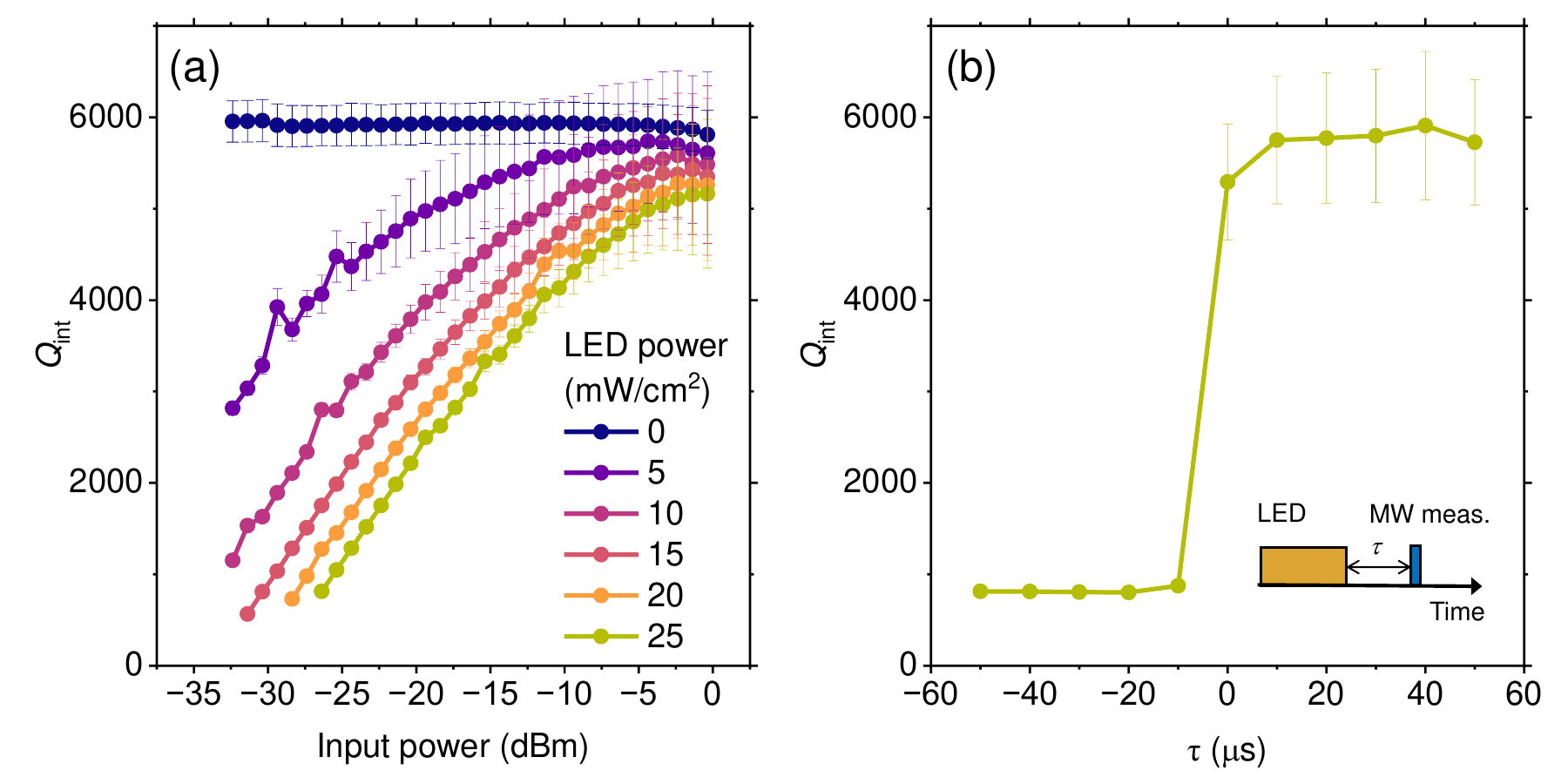}
    \caption{(a) The microwave input power dependence of $Q_{\rm int}$ under optical illumination with various powers. The measured sample is the bottom trap chip before assembly. The sample was irradiated with an LED light at a center wavelength of 308 nm and a bandwidth of 30 nm. The irradiation spot size was 3 mm in diameter, covering the entire resonator surface. (b) The measurement pulse delay time $\tau$ dependence of $Q_{\rm int}$ for an LED power of 25 mW/cm$^2$ and a microwave input power of -29 dBm. The inset shows the pulse sequence. The delay time $\tau$ is defined as the time delay between the LED illumination pulse and the microwave measurement pulse.}
    \label{fig:Fig9}
\end{figure}
In practical applications, the presence of laser beams during cooling and readout of ions can potentially diminish the performance of superconducting trap chips due to quasiparticle excitation within the superconducting film. In this section, we explore how the internal $Q$ factor varies with light exposure on a superconducting chip. Specifically, we utilized the bottom trap chip previously employed in the experiment. We measured $Q_{\rm int}$ as a function of the light-emitting diode (LED) power. The LED utilized in the experiment emitted light with a center wavelength of 308 nm and a bandwidth of 30 nm. The focused spot on the chip had a diameter of 3 mm, effectively covering the entire resonator surface. Figure \ref{fig:Fig9}(a) illustrates $Q_{\rm int}$ measured at 3 K as a function of the microwave input power under various continuous illumination powers. At low microwave input power, $Q_{\rm int}$ deteriorates rapidly with increasing illumination power density from 0 to 25 mW/cm$^2$. This degradation is expected to arise from quasiparticle excitation within the Nb film. Interestingly, the diminished $Q_{\rm int}$ rebounds to its original value at high microwave input power. This phenomenon, known as quasiparticle redistribution \cite{Visser2014, Budoyo2016, Fischer2023}, occurs when a microwave drive induces a redistribution of the nonequilibrium quasiparticle density $f(E)$, such that $f(E) - f(E+\hbar\omega_{\rm m})$ decreases with increasing input power. Consequently, this reduction diminishes the optically induced real part of the complex conductivity, thereby restoring $Q_{\rm int}$. Therefore, the impact of light exposure on the microwave performance is minimal.

We also examined how fast the degraded $Q_{\rm int}$ at low microwave input power recovers to the original value when the illumination is switched off. For this experiment, we employed square voltage pulses to modulate the LED output and trigger the single-frequency measurement of the VNA with a delay time $\tau$ from the moment the LED output was switched off (refer to the inset of figure \ref{fig:Fig9}(b)). Utilizing various delay times, we conducted microwave spectra measurements by sweeping the measurement frequency during the triggered single-frequency measurement. Subsequently, we re-estimated $Q_{\rm int}$ by fitting the obtained spectra. The $Q_{\rm int}$, degraded to $800$ during LED illumination, promptly rebounded to its original value of $5800$ within the measurement delay step of 10 $\mu$s upon switching off the LED output. Considering that the characteristic timescales for quasiparticle-phonon scatterings in Nb films are less than $\sim$ 150 ns \cite{PhysRevB.14.4854}, we expect a recovery timescale significantly faster than 10 $\mu$s. The recovery timescale of $< 10$ $\mu$s is much faster than the decoherence timescale of ion qubits and other ion-based operations. The fast recovery of the $Q$ factor ensures no degradation in the performance of microwave sideband cooling and microwave gates by halting laser beam emissions during these operations.

\section{Power efficient two-qubit gate scheme}
In this section, we propose a power-efficient two-qubit gate scheme by harnessing the high $Q_{\rm int}$ of superconducting resonators. Before exploring the power-efficient scheme, we first examine the power efficiency of the typical Mølmer–Sørensen (MS) gate \cite{PhysRevLett.82.1971}.

\begin{figure}
    \centering
    \includegraphics[width=1\linewidth]{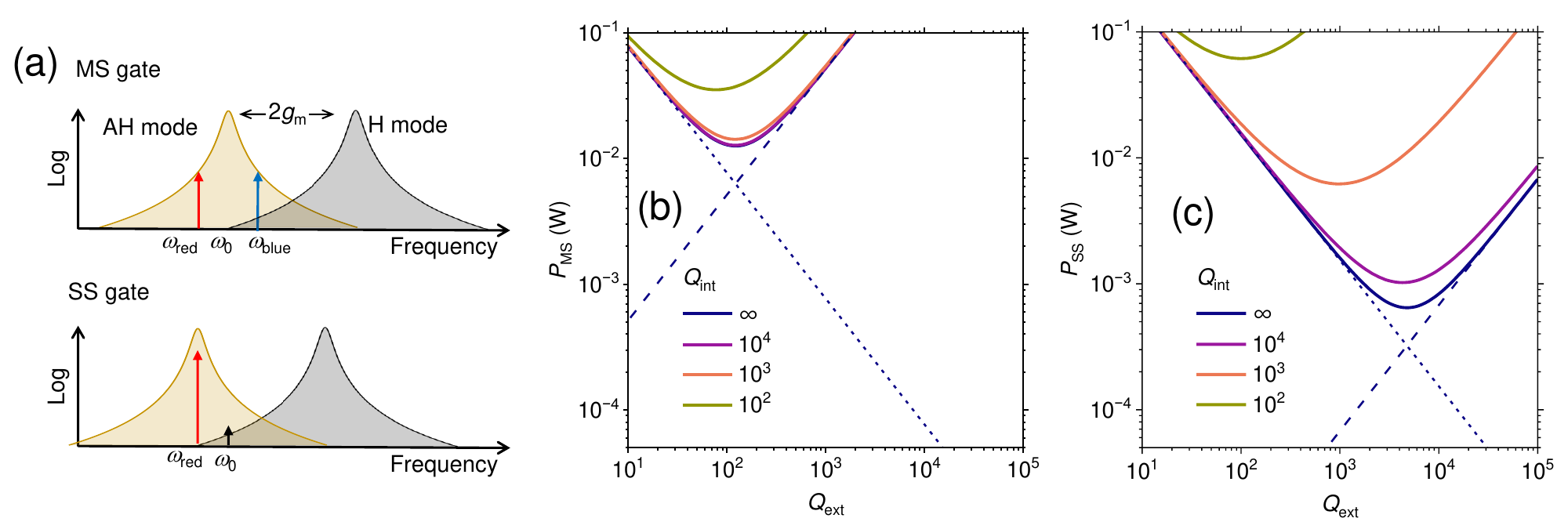}
    \caption{(a) Schematic diagrams of the MS and SS gate schemes in the frequency domain. The yellow and grey Lorentzian modes represent the AH- and H-modes split by $2g_{\rm m}$, respectively. For the MS gate scheme, red and blue sideband drives at $\omega_\mathrm{red}$ and $\omega_\mathrm{blue}$ interact with ions through the Lorentzian tails of the AH mode. For the SS gate scheme, a red sideband drive at $\omega_\mathrm{red}$ interacts with ions through resonant excitation of the AH mode. A carrier drive at $\omega_{0}$ excites a spin transition of ions through the Lorentzian tail of the H-mode. (b), (c) Required input powers $P_\mathrm{MS}$ and $P_\mathrm{SS}$ for the MS and SS gate schemes with various $Q_\mathrm{int}$ and $Q_\mathrm{ext}$. The sideband Rabi frequencies are assumed to be $\Omega_\mathrm{MS} = 2\pi\times 1$ kHz and $\Omega_\mathrm{SS} = 2\pi\times 2$ kHz. The dotted and dashed lines indicate the competing dependencies of $\propto 1/Q_\mathrm{ext}$ and $\propto Q_\mathrm{ext}$, respectively. }
    \label{fig:Fig10}
\end{figure}

Figure \ref{fig:Fig10}(a) illustrates a frequency diagram of the MS scheme, considering two Be$^{9+}$ ions interacting with the resonator modes of the 3D trap chip. The low- and high-frequency resonance modes represent the AH and H modes of the coupled resonators, respectively. The AH mode's center frequency resonates with a spin transition frequency $\omega_{0}$. While one can consider using various types of spin transitions for the gate, we assume the $\pi$-transition $\ket{F=2,m=0} \rightarrow \ket{F=1,m=0}$, which has a large magnetic transition moment and low-frequency sensitivity at low DC magnetic fields. The DC magnetic field assumed here lies within the xy plane and has an angle $\phi$ from the x-axis. The application of a DC magnetic field $B_0 = 1$ mT resolves the degeneracy among the hyperfine transition frequencies, producing a frequency separation of $\sim 7$ MHz. This field strength is much weaker than the critical magnetic field of Nb films \cite{Kwon2018, Wang2023}. Hence, any attendant reduction in the $Q$ factor and $I_{\rm max}$ attributable to this supplementary DC magnetic field is expected to be negligible. The $\pi$-transition interacts with a microwave magnetic component parallel to its magnetic transition moment, as discussed in section 3.

To realize the MS-type interaction, we employ red and blue sideband drives with small frequency detunings $\pm\delta$ from motional sidebands of the spin transition. The red and blue drive frequencies are described as $\omega_{\rm red} = \omega_{0} - \omega_{\rm rock} - \delta$ and $\omega_{\rm blue} = \omega_{0} + \omega_{\rm rock} + \delta$, where $\omega_{\rm rock}$ denotes the rocking mode frequency of the ions. We assume these drives only generate magnetic field gradients at the ion positions without exciting magnetic field components. This can be achieved by selectively driving the AH mode using inputs to each resonator with proper phase offsets.

To determine the necessary power for the MS gate to achieve a specific sideband coupling $\Omega_{\rm M}$, we solved equations (\ref{eq:4}) and (\ref{eq:5}) to obtain the relation between an input power $P_{\rm MS}$ and a resonator photon number needed to achieve the desired $\Omega_{\rm M}$. The required power for the MS gate is given by
\begin{eqnarray} \label{eq:6}
P_{\rm MS} = \hbar(4\omega_{\rm rock}^2+\frac{\omega_r^2}{Q_{\rm total}^2})Q_{\rm ext}(\frac{\hbar\Omega_{\rm M}}{-\mu_{\parallel}\frac{\partial{B^{\rm AH0}_{x}(0, \theta_{\rm HF})}}{\partial{r}}|b_j|q^0 \cos\phi})^2,
\end{eqnarray}
Here, we assume that the top and bottom resonators are identical. Additionally, we assume $\omega_r \gg g_{\rm m}, \omega_{\rm rock} \gg \delta$ and the red and blue sideband Rabi frequencies $\Omega_{\rm M}$ are the same. Other parameters include $\mu_{\parallel}$, $\theta_{\rm HF}$, $\phi$, $b_j$ (rocking mode amplitude of $j$th ion), $q^0_j$ (zero point motion of $j$th ion), and $B^{\rm AH0}_x$ (single-photon magnetic field of the AH mode derived from a COMSOL simulation). The last fraction in equation (\ref{eq:6}) represents a resonator photon number needed to achieve a certain $\Omega_{\rm M}$.

Figure \ref{fig:Fig10}(b) presents the values of $P_{\rm MS}$ required to achieve $\Omega_{\rm {M}} = 2\pi\times1$ kHz for various $Q_{\rm {ext}}$ and $Q_{\rm {int}}$. The parameters used in this calculation are summarized in Table 1. As $Q_{\rm {int}}$ decreases, the required power $P_{\rm {MS}}$ decreases, reaching a minimum of $P_{\rm {MS}} \sim 14$ mW when $Q_{\rm {int}} > 1000$. This minimum is determined by the two competing dependencies shown as the dashed and dotted lines. For low $Q_{\rm {ext}}$ ($(2\omega_{\rm {rock}}/\kappa_{\rm {ext}})^2 \ll 1$), $P_{\rm {MS}}$ decreases following the dependency $P_{\rm {MS}} \propto 1/Q_{\rm {ext}}$ (dotted line). In this region, a higher $Q$ factor of the AH mode allows for a larger gain of resonator photon number from the input. Conversely, for higher $Q_{\rm {ext}}$ ($(2\omega_{\rm {rock}}/\kappa_{\rm {ext}})^2 > 1$), $P_{\rm {MS}}$ increases according to the dependency $P_{\rm {MS}} \propto Q_{\rm {ext}}$ (dashed line). This is due to limited input coupling to the AH mode, as the drive frequencies $\omega_{\rm {red}}$ and $\omega_{\rm {blue}}$ fall out of the AH mode bandwidth.

\begin{table}
\caption{\label{math-tab2}Parameters used in the calculations for Figures 10(b) and 10(c).}
\begin{tabular*}{\textwidth}{@{}l*{15}{@{\extracolsep{0pt plus 12pt}}l}}
\br
Parameters&Description&Values\\
\mr
$\mu_{\parallel}$&Magnetic transition moment&$-9.28\times10^{-24}$ J/T\\
$\partial{B^{\rm AH0}_x(0, \theta_{\rm HF})}/\partial{r}$&AH mode single-photon field gradient&$1.2\times10^{-6}$ T/m\\
$B^{\rm H0}_x$&H mode single-photon field&$5.9\times10^{-11}$ T\\
$|b_j|$&Rocking mode amplitude&$1/\sqrt{2}$\\
$q^0_j$&Zero point motion of $j$th ion&$9.30\times10^{-9}$ m\\
$\phi$&in-plane DC magnetic field angle&$45\degree$\\
$\theta_{\rm HF}$&HF motional mode angle&$36\degree$\\
$g_{\rm m}$&Resonator normal mode splitting&$2\pi\times30$ MHz\\
$\omega_{\rm rock}$&Rocking mode frequency&$2\pi\times4.4$ MHz\\
$\omega_r$&Bare resonator frequency&$2\pi\times1.074$ GHz\\
$\Omega_{\rm M}$&Mølmer–Sørensen sideband Rabi frequency&$2\pi\times1$ kHz\\
$\Omega_{\rm S}$&Single-sideband Rabi frequency&$2\pi\times2$ kHz\\
$\Omega_{\rm C}/\Omega_{\rm S}$&Carrier to sideband Rabi ratio&$15$\\
\br
\end{tabular*}
\end{table}

To enhance the power efficiency of two-qubit gates by overcoming the limitations posed by the resonance bandwidth, we propose a single-sideband (SS) gate scheme. The original idea of the SS gate scheme was proposed in the context of cavity QED systems \cite{Zheng2002, Solano2003} and later proposed and demonstrated in a trapped ion system \cite{Bermudez2012, Tan2013}. The bottom row of figure \ref{fig:Fig10}(a) illustrates the frequency diagram of the gate scheme. In this scheme, the center frequency of the AH mode is resonant with $\omega_{\rm {red}}$ to efficiently generate the magnetic field gradient. Conversely, the carrier drive at frequency $\omega_0$ rotates the spin state by inducing a homogeneous magnetic field through the Lorentzian tail of the H mode. The AH and H modes can be selectively excited by driving each resonator with proper phase offsets. The corresponding SS Hamiltonian in the interaction picture is given by:
\begin{equation} \label{eq:7}
H_{\rm SS} = \sum_{j=1,2}\{\frac{\hbar \Omega^j_{\rm C}}{2}(\sigma^+_j + \sigma^-_j) + \frac{\hbar\Omega^j_{\rm S}}{2}(e^{-i \delta t} a^\dag \sigma^-_j + e^{i \delta t} a \sigma^+_j)\},
\end{equation}
In this expression, $\sigma^{+(-)}j$ refers to the creation (annihilation) operator of the spin-up and spin-down states of the $j$th ion, denoted as $\ket{\uparrow_j}$ and $\ket{\downarrow_j}$, respectively. Similarly, the creation (annihilation) operator of the rocking mode is represented by $a^\dag$ ($a$), and $\Omega^j_{\rm C}$ ($\Omega^j_{\rm S}$) is the spin flip-flop (sideband) Rabi frequencies of the $j$th ion. Each coupling strength is described as:
\begin{eqnarray} \label{eq:8}
\Omega^j_{\rm C} = -\frac{\mu_{\parallel}}{\hbar} B^{\rm H}_x \cos{\phi}\\
\label{eq:9}
\Omega^j_{\rm S} = -\frac{\mu_{\parallel}}{\hbar}\frac{\partial{B^{\rm AH}_x}(0, \theta_{\rm HF})}{\partial{r}}b_jq^0_j \cos\phi,
\end{eqnarray}
where the x-axis components of the magnetic fields excited through the H and AH modes are denoted as $B^{\rm H}_x$ and $B^{\rm AH}_x(r, \theta)$, respectively. Hereafter, we assume $\Omega^j_{\rm C} = \Omega_{\rm C}$ and $|\Omega^j_{\rm S}| = \Omega_{\rm S} = -\frac{\mu_{\parallel}}{\hbar}\frac{\partial{B^{\rm AH}_x(0, \theta_{\rm HF})}}{\partial{r}}|b_j|q^0_j$. After moving onto a frame rotating at $\Omega_{\rm C}$ and assuming $\Omega_{\rm C} \gg \Omega_{\rm S}, \delta$, one can obtain the Hamiltonian:
\begin{equation} \label{eq:10}
H_{\rm SS}^{\prime} = \frac{\hbar\Omega_{\rm S}}{2}(e^{-i \delta t} a^{\dag} + e^{i \delta t} a) S_x.
\end{equation}
Here, $S_x = \frac{1}{2} \sum_{j=1,2} sgn(b_j)\left(\sigma^+_j + \sigma^-_j\right)$.
A red sideband drive with a strong carrier drive induces a spin-dependent force on the ions. By applying the drives for a time $\tau_{\rm S} = 2\pi/\delta = 2\pi/\Omega_{\rm S}$, one can produce a Bell state $\ket{\uparrow\downarrow} - i\ket{\uparrow\downarrow}$ from an initial state $\ket{\uparrow\downarrow}$. However, neglected terms in equation (\ref{eq:7}) under the assumption of $\Omega_{\rm C} \gg \Omega_{\rm S}, \delta$ lead to a small Bell state infidelity at a finite $\Omega_{\rm C}$. This infidelity is attributed to the Stark shift of states $\ket{+_j} = \ket{\uparrow_j}+i\ket{\downarrow_j}$ and $\ket{-_j} = \ket{\uparrow_j}-i\ket{\downarrow_j}$, and can be reduced by increasing $\Omega_{\rm C}/\Omega_{\rm S}$. By numerically solving equation (\ref{eq:7}), we confirm that the additional infidelity compared with the MS gate is suppressed to be $7\times10^{-4}$ when $\Omega_{\rm C}/\Omega_{\rm S} = 15$. At this Rabi ratio, an additional microwave resonator current to obtain $\Omega_{\rm C}$ is much smaller than that for $\Omega_{\rm S}$. Hence, the amount of additional current for the carrier drive is negligible with respect to the critical current.

We explored the necessary input powers for the SS gate scheme by solving equations (\ref{eq:4}) and (\ref{eq:5}), yielding the resonator parameter dependencies of the input powers for the carrier drive ($P_{\rm C}$) and red sideband drive ($P_{\rm R}$):
\begin{eqnarray} \label{eq:11}
P_{\rm C} = \frac{\hbar(4(2 g_{\rm m} - \omega_{\rm rock})^2 + (\frac{\omega_r}{Q_{\rm total}})^2)Q_{\rm ext}}{2}(\frac{\hbar \Omega_{\rm C}}{-\mu_{\parallel} B^{\rm H0}_x\cos{\phi}})^2,\\
\label{eq:12}
P_{\rm R} = \frac{\hbar(\frac{\omega_r}{Q_{\rm total}})^2Q_{\rm ext}}{2}(\frac{\hbar\Omega_{\rm S}}{-\mu_{\parallel}\frac{\partial{B^{\rm AH0}_x(0, \theta_{\rm HF})}}{\partial{r}}|b_j|q^0_j \cos\phi})^2.
\end{eqnarray}
Figure \ref{fig:Fig10}(c) illustrates the total input power $P_{\rm SS} = P_{\rm C} + P_{\rm R}$ required to achieve $\Omega_{\rm S} = 2\pi\times2$ kHz and $\Omega_{\rm C}/\Omega_{\rm S} = 15$ for various $Q_{\rm ext}$ and $Q_{\rm int}$. We designated $\Omega_{\rm S} = 2\Omega_{\rm M}$ to ensure that the gate time of the SS gate ($\tau_{\rm S} = 2\pi/\delta = 2\pi/\Omega_{\rm S}$) equals $\tau_{\rm S} = \tau_{\rm M}$. Other parameters employed in this calculation are summarized in Table 1. When $Q_{\rm int} \geq 10^3$, the required $P_{\rm SS}$ diminishes to a substantially smaller value than that for the MS gate, reaching the minimum of $P_{\rm SS} = 0.65$ mW. For sufficiently large $Q_{\rm int} \gg Q_{\rm ext}$, the minimum $P_{\rm SS}$ is determined by competing dependencies of $P_{\rm C}$ and $P_{\rm R}$, as depicted with the dashed and dotted lines, respectively. For the low $Q_{\rm ext}$ side, $P_{\rm R}$ decreases following the dependency $P_{\rm R} \propto 1/Q_{\rm {ext}}$ (dotted line). Since the red sideband drive resonates with the AH mode, a higher $Q_{\rm ext}$ results in reduced $P_{\rm R}$ owing to a larger gain of AH mode photon number. On the other hand, for the high $Q_{\rm ext}$ side, $P_{\rm C}$ increases according to the dependency $P_{\rm C} \propto Q_{\rm {ext}}$ (dashed line). This arises from the inefficiency of the carrier drive coupling to the H mode with increasing $Q_{\rm ext}$ due to the limited H mode bandwidth. Unlike the MS scheme, however, the large single-photon coupling $-\mu_{\parallel} B^0_x \cos{\phi}$ of the carrier transition requires a much smaller input power than the sideband transitions to achieve a certain coupling strength. This reduces the power limit on the high $Q_{\rm ext}$ side compared to the MS gate, allowing us to achieve $P_{\rm SS} = 1.0$ mW at $Q_{\rm int} = 10^4$. This represents a reduction of three orders of magnitude compared to the power requirement for the MS gate with normal metallic surface trap chips.

\section{Prospects}
\subsection{Further improvements on the maximum currents}
We have demonstrated the substantial $I_{\rm max}$ achievable with high $Q^{\rm m}_{\rm int}$ using both superconducting 2D and 3D trap chips. However, in the case of the 3D trap chip, the attainable $Q^{\rm m}_{\rm int}$ is limited by the loss originating from the silver paste layers utilized for electrical connections to the top chip. The relatively low $Q^{\rm m}_{\rm int}$ suggests that the obtained $I_{\rm max}$ is also restricted, likely due to excessive heat generation from the Joule loss. To address these challenges, substituting low-loss connections, such as aluminum or gold wire bondings, for these electrical connections would be an effective strategy. We anticipate achieving similar $I_{\rm max}$ and $Q^{\rm m}_{\rm int}$ values as those observed in the single LC resonator.

To further enhance $I_{\rm max}$ while preserving high $Q^{\rm m}_{\rm int}$, one can pursue several strategies. One approach involves mitigating $B_{\rm s}$ at the film corners, as this concentrated field disrupts superconductivity, thereby limiting $I_{\rm max}$. As discussed in Section 2, increasing the film thickness from 1 to 5 $\mu$m reduces $B_{\rm s}$ by a factor of 1.4. This enhancement should project to the increase in $I_{\rm max}$ by the same factor. To increase the thickness, one has to be careful of accumulated stress on sputtered Nb films since strong tensile or compressive stress causes cracks in the films or deteriorates the quality of the substrate-film interface. One can suppress the film stress by performing a multilayer sputtering process, where a cooling time inserted between each sputtering layer allows adatoms to rearrange on the surface and relax the film stress \cite{Palmieri:SRF2017-TUYBA03}.           

Integrating a nanometric multilayer structure on a thick Nb film is also an effective option to reach a higher magnetic field \cite{Kubo2017}. The multilayer structure comprises alternating insulators and thin superconducting films with a thickness smaller than the magnetic penetration depth of Nb. The material of the thin superconducting layer must have a higher $H_{\rm c}$ than that of Nb. Such thin film multilayer structure acts as a barrier that prevents the magnetic vortices from penetrating into the thick Nb layer, while also suppressing the microwave loss of the vortices nucleated in the thin film layers. A multilayer of (MgO/NbN)$_4$ on a Nb film has demonstrated an increase in the maximum field by ten-fold compared with a bare Nb film \cite{Antoine2013}. 

Another approach involves enhancing the low RRR in the Nb film, as the heating of impurities within the film is likely the primary cause of the breakdown. Conventionally, magnetron-sputtered Nb films exhibit relatively low RRR values ($\sim$ 10) due to the incorporation of impurity atoms within the film during deposition \cite{Marie2016}. Over the years, numerous growth conditions and techniques have been explored. Notably, methods like energetic condensation growth, such as vacuum arc deposition and coaxial energetic deposition, have enabled the attainment of RRR values in the several hundred ranges, comparable to those of pure bulk Nb  \cite{Marie2016}.

\subsection{Resonator frequency tunability for the QCCD architecture}
For actual trapped-ion quantum computing, integrating the 3D trap structure into the quantum charge-coupled device (QCCD) architecture emerges as a promising avenue \cite{Kielpinski2002, Pino2021}. In this architecture, necessary operations are performed at spatially different zones interconnected by shuttling ions \cite{Blakestad2009, Walther2012}. Each zone serves a specific purpose, such as gate operations, state readout, and ion storage. Conceptually, all zones can adopt the flip-chip geometry, incorporating superconducting DC and RF electrodes alongside zone-specific components, such as microwave resonators for gate zones and superconducting photon detectors for readout zones.

Efficient implementation of the power-efficient two-qubit gate scheme across gate zones within the QCCD architecture hinges upon achieving frequency matching between the motional red sideband of ions and the AH mode resonances of all resonators over multiple gate zones. Currently, the resonance frequencies of our fabricated resonators exhibit a deviation of approximately 10 MHz within the same batch. One potential strategy to fulfill the frequency-matching condition involves post-fabrication tuning of the resonance frequencies of each resonator while spin transition frequency is fixed at a specific value by applying a weak DC magnetic field.

Traditionally, tuning the resonance frequencies of high-$Q$ superconducting resonators involves manipulating the tunable kinetic inductance facilitated by Josephson junctions \cite{Beltran2007, Strickland2023}. However, this approach encounters challenges due to the incompatibility of Josephson junctions with high-power microwave drives. Alternatively, adjustments to the resonator capacitance offer a viable solution. Integration of a superconducting vacuum gap structure into each resonator enables capacitance adjustment by tuning the gap distance with a bias voltage between the gap. Notably, a superconducting LC resonator featuring a vacuum gap capacitor has demonstrated a $Q$ factor exceeding $10^4$ and frequency tunability spanning several tens of MHz, effectively accommodating variations in resonance frequencies \cite{8990929, Jiang2020}.

Beyond two-qubit gate operations, efficient execution of individual qubit manipulation within the same gate zone holds another importance for the QCCD architecture. Unlike laser beams, confining microwave fields to individual ions within the same gate zone poses a significant challenge. Instead, harnessing a magnetic field gradient to drive a micromotion sideband offers a viable approach to accessing individual ions with minimal crosstalk errors \cite{Warring2013}. Since the micromotion sideband frequency is far detuned from the red sideband frequency beyond the AH mode bandwidth, tuning the AH mode frequency between them depending on two-qubit or individual qubit manipulation becomes necessary. Consequently, the aforementioned frequency tunability of microwave resonators emerges as a critical requirement in this attempt.

\section{Summary}
For low-loss and power-efficient microwave operations, we demonstrated a superconducting surface trap chip that features a large microwave current and a high $Q$ factor simultaneously. To this end, we first designed and fabricated 2D surface trap chips with electrodes made of superconducting Nb. Thanks to the large critical field of Nb films, we achieved $I_{\rm max} = 0.74$ A with $Q^{\rm m}_{\rm int} = 6000$, resulting in $\partial{B_z}/\partial{x} = 28$ T/m for a 2D surface trap geometry. To generate a larger magnetic field gradient, we also developed a superconducting 3D trap chip. The microwave LC superconducting resonators incorporated in the trap chip demonstrate $I_{\rm max} = 0.56$ A, generating $\partial B^{\rm AH}_x(0, \theta_{\rm HF})/\partial r = 42$ T/m, and $Q^{\rm m}_{\rm int} = 1600$ at 5 K. The fabricated 3D superconducting trap chip can handle a comparable amount of current and has an order of magnitude high $Q$ factor compared with normal metal trap chips. By combining design and material improvements, the performance of superconducting surface trap chips could be significantly elevated. This combined effort potentially leads to achieving $I_{\rm max} > 1$ A, resulting in $\partial B^{\rm AH}_x(0, \theta_{\rm HF})/\partial r > 100$ T/m with $Q^{\rm m}_{\rm int} > 10^4$ at 5 K.

Alongside the large $I_{\rm max}$ and high $Q^{\rm m}_{\rm int}$, we have also demonstrated a remarkable resilience of $Q^{\rm m}_{\rm int}$ to optical illumination. When a superconducting resonator is subjected to optical illumination under a weak microwave drive, $Q_{\rm int}$ experiences degradation but promptly recovers within a time scale of less than 10 $\mu$s after the illumination ceases. Since laser beams can be deactivated during microwave operations, such as sideband cooling and gate sequences, the rapid recovery of $Q_{\rm int}$ facilitates microwave operations without compromising performance. Furthermore, even in the presence of optical illumination, a strong microwave drive effectively restores $Q_{\rm int}$ to nearly its original value by quasiparticle redistribution. This observation implies that any stray light from concurrently operated laser operations in neighboring zones within the QCCD architecture does not interfere with microwave gate operations in the targeted zone.

Finally, we have demonstrated that the required input power for a two-qubit gate can be as low as 1.0 mW, with $Q_{\rm int} = 10^4$ and a gate time of 1 ms. To achieve this, we adopted the SS gate scheme, which employs a red sideband drive and a carrier drive instead of the conventional MS gate. Leveraging the high $Q_{\rm int}$ of superconducting resonators, the SS gate enables two-qubit gates with orders of magnitude lower power consumption compared to normal metallic trap chips.

In conclusion, the findings presented in this paper highlight the potential of superconducting surface trap chips as a promising solution for realizing low-loss and power-efficient microwave-driven trapped ion systems. Future work will employ these superconducting surface trap chips to trap ions and perform quantum gate operations with comparable or faster time scales, all while maintaining significantly lower microwave power consumption and dissipation compared to conventional normal metal surface trap systems.

\ack
We acknowledge Hirotaka Terai for providing Nb films and appreciate insightful discussions about Nb films with Kensei Umemori, Tatsumi Nitta, Toshiaki Inada, Hayato Ito, and Takayuki Kubo. We also thank Yasunobu Nakamura and the superconducting quantum electronics research team for their support in device fabrication. This work has been supported by the JST Moonshot R$\&$D Program (Grant No. JPMJMS2063) and JST-PRESTO (Grant No. JPMJPR2258).

%\section*{Data availability statement}
%The data that support the findings of this study are available upon reasonable request from the authors.

%\section*{Conflict of interest}
%The authors declare no competing financial or non-financial interests.

\section*{References}
\bibliographystyle{unsrt}
\bibliography{mybib}

\end{document}